%% file: TCOM_TPS-20-1290_rev.tex
\documentclass[journal,draftcls,onecolumn,12pt,twoside]{IEEEtranTCOM}
\normalsize
\usepackage{amsopn, amsmath, fancybox, epsfig, amssymb, color, bm, dsfont , latexsym, graphicx,
accents, subfigure, algorithm, algpseudocode, multirow, pifont, caption, wasysym, stmaryrd, bbm, tablefootnote,xurl}

\newtheorem{property}{Property}

\newcommand{\squeezeup}{\vspace*{-0cm}}
\newcommand{\squeezeupAlgo}{\vspace{-0.15cm}}
\newcommand{\BSGN}[1]{\text{bsgn}#1}

\begin{document}

\title{Belief-Propagation Decoding of LDPC Codes with Variable Node--Centric Dynamic Schedules}
\author{Tofar~C.-Y.~Chang,~\IEEEmembership{Member,~IEEE},
Pin-Han Wang,~
Jian-Jia Weng,~\IEEEmembership{Member,~IEEE},
I-Hsiang Lee,~
and~Yu~T.~Su,~\IEEEmembership{Life Senior Member,~IEEE}
}
\maketitle
\input{Abstract.tex}
\input{Sec1_Intro.tex}

\input{Sec2_Review.tex}
\input{Sec3_PDRBP.tex}
\input{Sec4_LMDRBP.tex}
\input{Sec5_Result.tex}

\input{Sec6_MultiEdge.tex}

\input{Sec7.tex}
\appendices
\renewcommand{\thesection}{Appendix \Alph{section}}
\renewcommand{\theequation}{\Alph{section}.\arabic{equation}}
\setcounter{equation}{0}
\input{Appendix_0214.tex}
\input{ref.tex}

\end{document}

%% file: Abstract.tex
\begin{abstract}
Belief propagation (BP) decoding of low-density parity-check (LDPC) codes with various dynamic decoding schedules have been proposed to
improve the efficiency of the conventional flooding schedule. As the ultimate goal of an ideal LDPC code decoder is to have correct
bit decisions, a dynamic decoding schedule should be variable node (VN)-centric and be able to find the VNs with probable incorrect decisions
and having a good chance to be corrected if chosen for update. We propose a novel and effective metric called conditional innovation (CI)
which serves this design goal well. To make the most of dynamic scheduling which produces high-reliability bit decisions, we limit our
search for the candidate VNs to those related to the latest updated nodes only.

Based on the CI metric and the new search guideline separately or in combination, we develop several highly efficient decoding schedules.
To reduce decoding latency, we introduce multi-edge updating versions which offer extra latency-performance tradeoffs. Numerical results
show that both single-edge and multi-edge algorithms provide better decoding performance against most dynamic schedules and the CI-based
algorithms are particularly impressive at the first few decoding iterations.

\end{abstract}


%% file: Sec1_Intro.tex
\section{Introduction}\label{section:intro}
Low-density parity-check (LDPC) codes are known to provide near-capacity performance when the belief propagation
(BP) algorithm is utilized for decoding \cite{LDPC}. These codes have been used in many applications such as
deep-space network, disk storage, satellite communications and have adopted by several wireless communication standards,
e.g., IEEE 802.11 (WiFi) \cite{WiFi} and 5G New Radio (NR) \cite{NR}.

The conventional BP algorithm performs message-passing on the code graph based on the \emph{flooding} scheduling:
the variable-to-check (V2C) messages sent from all variable nodes (VNs) to the linked check nodes (CNs) are updated
and propagated simultaneously, so are the check-to-variable (C2V) messages. However, such a fully-parallel decoding
schedule often requires many iterations to converge and necessitates complicated interconnections and large memory
for hardware implementation. Therefore, sequential and semi-sequential decoding schedules have been proposed for
improving the convergence speed and/or reducing the implementation complexity \cite{LBP}-\cite{IDS19b}; some even
provide improved converged error rate performance. The non-flooding schedules are generally categorized into two
classes--the ordered schedules and the dynamic schedules. The former class is also referred to as the standard
sequential scheduling (SSS) strategies. The SSS-based BP decoders include the layered BP (LBP) \cite{LBP}, shuffled
BP \cite{SBP}, and their variants \cite{SBP_v1}, \cite{LBP_v1}. They converge at least twice faster than the
conventional BP decoder and require less processing time and simpler hardware implementation \cite{LDC1},
\cite{LDC2}.

The dynamic schedules modify the message-passing order based on newest available information. The informed dynamic
scheduling (IDS) strategies form a popular subclass of the dynamic schedules. It makes an element-wise comparison
of two sets we refer to as the current and precomputed message sets. The former set can be the set of the C2V, V2C
messages sent or VNs' total log-likelihood ratios (LLRs) computed in the last update (messages or LLRs may be updated
in different time instants) and the elements of the latter set are the corresponding values if updated. These messages
are functions of the channel values associated with each coded bits
(or VNs) which vary from a codeword to another and the messages collected from connecting VNs or CNs which vary with
each update. Hence, a proper dynamic schedule which adjusts the message-passing order according to these two message sets
may yield faster convergence speed and lower error rate. The IDS strategies forward only the best precomputed message(s)
according to a certain metric. In the residual BP (RBP) algorithm \cite{RBP}, the current and precomputed C2V message
sets are adopted, and element-wise differences between these two sets are called \emph{residuals}. The RBP algorithm
passes only the C2V message for the edge with the maximum residual among all code graph edges. It yields better convergence
speed in comparison with the SSS and flooding scheduling based BP decoders but suffers from inferior converged error
rate performance. The degraded performance is due in part to the greedy behavior that the decoder may keep updating
only a small group of edges \cite{RBP}. Hence, several algorithms were proposed to prevent such a greedy event
\cite{RBP}-\cite{IDS19a}. Other recent works have put more emphasis on improving the RBP algorithm's error rate performance.
In \cite{IDS11a}-\cite{IDS16a}, the message updating priority is mainly based on the VN decisions' stability while
the method proposed in \cite{IDS19b} makes use of the VNs' total LLR difference and increases the chance of an unreliable
VN to obtain the information originated from some reliable VNs. All these works have tried to improve the error rate
performance and/or convergence speed of an LDPC code decoding by selecting the optimal edge(s) for updating.  Healy {\it et al.}
\cite{IDS18b} simplified the precomputing task by considering, for a CN, only two connecting edges with the smallest V2C
magnitudes and selecting the one with the larger C2V residuals as this CN's candidate edge. Among all candidate edges,
the one with the largest C2V residual is chosen and the corresponding C2V message is propagated. Wang {\it et al.} \cite{LBP20a}
developed a fixed LBP decoding schedule which arranges the C2V message-passing order according to the least-punctured
and highest-degree principle. The authors also proposed a dynamic LBP schedule which slightly outperforms the fixed one.

As the ultimate decoding goal is to have correct VN decisions, an effective schedule should be VN-centric and focus
on accurately identifying the incorrect or unreliable VN decisions during decoding. It should give higher updating priority
to those which are most likely to be corrected. The VN reliability measurements, e.g., decision reversion \cite{IDS11a}-\cite{IDS16a},
the unsatisfied CN number \cite{IDS15a}, \cite{IDS19b} and the change of VNs' total LLRs \cite{IDS15a}-\cite{IDS19b} were
used implicitly to identify incorrect bit decisions and the unreliable VNs were given higher priority for update. On the
other hand, BP decoding is usually performed in LLR domain for computational simplicity and for the fact that the likelihood
can be recovered from its LLR value. However, the likelihood is a true bit decision reliability indicator and the change of
a VN's LLR is not linearly proportional to the likelihood or conditional probability variation. Thus there is clearly a need
to develop a new metric to suit the purpose of correcting the most proper erroneous or most unreliable decisions.
Moreover, many IDS decoders need to globally search for an edge/node for update and a reduction of the search range is
necessary.

In this paper, we propose an efficient metric, which we call \emph{conditional innovation (CI)}, to estimate
the potential likelihood improvement of a VN. CI is defined as the difference of a VN's current and precomputed
conditional posterior probabilities. We show that it also reflects the reliability or correctness of the corresponding
VN decision. We further verify that a larger CI not only implies that the corresponding VN decision is more likely to
be erroneous but also have a higher probability of being corrected if updated. The need of search range reduction and the
intuition that the latest updated messages tend to be more trustworthy than the others motivate us to introduce an updating
strategy that limits the next update candidates to those VNs which can be reached by the latest updated VNs in just two hops.
We demonstrate that the proposed strategy does enhance the reliability of the propagated messages and narrow the candidate
selection range.

Making use of these two concepts separately or in combination, we derive several efficient scheduling algorithms.
In particular, by adopting the CI as the reliability measure in the scheduling strategies, we develop a CI based RBP
(CIRBP) algorithm which is able to identify and correct most erroneous decisions in the first few iterations.
Therefore, our CIRBP algorithm provides excellent error rate performance at the early decoding stage and is shown
to outperform the existing IDS-based BP decoders. We also propose the \emph{latest-message-driven (LMD)} strategy
which uses the latest updated C2V messages to determine the next updated VN. We call the BP decoders which
employ the LMD strategy as the LMD-based RBP (LMDRBP) algorithms. The LMDRBP algorithms not only use the newest
updated messages in selecting the next updated VN but allow these newest messages to be passed with higher priority.
Simulation results indicate that the LMDRBP algorithms are able to surpass the existing RBP decoders for most cases
while requiring less or the same computation efforts in selecting the updated node or edge. Combining both LMD
strategy and CI metric, the resulting LMD-CIRBP algorithm obtains very impressive decoding gain, especially at
the early iterations, at the cost of moderate complexity increase.

As the edge-wise updating strategies presented in \cite{RBP}-\cite{IDS19a} are performed in a fully-serial manner,
i.e., only one of the precomputed messages is propagated in each update, they entail long decoding delays.
As far as the decoding delay is concerned, those which adopt multi-edge updating (\cite{RBP}, \cite{IDS11a}-\cite{IDS19b})
and propagate more than one messages per update clock are more practical. The increased parallelism reduces the decoding
latency but may cost performance loss. We develop multi-edge updating versions of our CIRBP and LMD-CIRBP algorithms
with the degree of parallelism as an adjustable parameter to provide performance-latency tradeoff. Experimental results
show that, using a judicial chosen parallelism, our multi-edge LMD-CIRBP algorithm achieves much reduced latency per
iteration with little or no performance loss with respect to its single-edge counterpart.

The rest of this paper is organized as follows. In Sec. \ref{section:review}, we give a brief review of
known RBP algorithms and the corresponding IDS strategies used. The properties of the proposed CI metric are analyzed
and the CIRBP algorithm is presented in Sec. \ref{section:PD}. In Sec. \ref{section:LMD}, we discuss the LMD scheduling
strategy and present the LMDRBP and LMD-CIRBP algorithms. The numerical results and complexity analysis of our decoders
are provided in Sec. \ref{section:simulation}. In Sec. \ref{section:multiEdge}, we introduce the multi-edge
updating versions of the CIRBP and LMD-CIRBP algorithms and give related simulation results. Finally, we draw concluding
remarks in Sec. \ref{section:conclusion}.

%% file: Sec2_Review.tex
\section{Preliminaries}\label{section:review}
\subsection{Message Updating for RBP Decoding}\label{subsection_BP}
A binary ($N$, $K$) LDPC code $\mathcal{C}$ of rate $R = K/N$ is characterized by an $M \times N$ parity-check matrix
$\bm{H}=[h_{mn}]$, where the entry $h_{mn}$ determines if the $n$th VN $v_n$ and the $m$th CN $c_m$ on the associated
bipartite code graph is connected. For a coded BPSK system, a binary codeword $\bm{u}=(u_0,u_1,\cdots,u_{N-1}), u_n \in
\{0,1\}$ is modulated to the sequence $\bm{x}=(x_0,x_1,\cdots,x_{N-1})$, where $x_n = 1-2 u_n$ for $0\leq n < N$,
and then transmitted over an AWGN channel. The corresponding received noisy sequence and tentative decoded decision
vector are respectively denoted by $\bm{y}=(y_0,y_1,\cdots,y_{N-1})$ and $\hat{\bm{u}}=(\hat{u}_0,\hat{u}_1,\cdots,\hat{u}_{N-1})$,
where $y_n = x_n+w_n$ and $w_n,~ 0\leq n < N$, are 
i.i.d. zero-mean AWGN with variance $\sigma^2$.

Let $L_{m\rightarrow n}^{\text{C}}$ be the C2V message from $c_m$ to $v_n$ in a BP-based decoder, $L_{n\rightarrow m}^{\text{V}}$
be the V2C message from $v_n$ to $c_m$, and $L_n$ be the total LLR of $v_n$. For all $m,n$ such that
$h_{mn}=1$, $L_{m\rightarrow n}^{\text{C}}$ and $L_{n\rightarrow m}^{\text{V}}$ are initialized as $0$ and $2y_n/\sigma^2$,
respectively. We denote by $\mathcal{M}(n)=\{m|h_{mn}=1\}$ the index set of CNs connected to $v_n$ and by $\mathcal{N}(m)=
\{n|h_{mn}=1\}$ the index set of VNs linked to $c_m$ on the associated code graph.
We further define $\mathcal{M}(n)\setminus m$ and $\mathcal{N}(m)\setminus n$ respectively as the set $\mathcal{M}(n)$ with $m$ excluded and the set $\mathcal{N}(m)$ with $n$ excluded.
For the BP decoding algorithm, the V2C messages sent from $v_n$ to $c_m$,
$m\in \mathcal{M}(n)$, are calculated by
\begin{eqnarray}\label{eqn:V2C}
  L_{n\rightarrow m}^{\text{V}} = \frac{2y_n}{\sigma^2} + \sum_{m'\in\mathcal{M}(n)\setminus m} L_{m'\rightarrow n}^{\text{C}},
\end{eqnarray}
and the C2V message sent from $c_m$ to $v_n$, $n\in \mathcal{N}(m)$ are updated by
\begin{eqnarray}\label{eqn:C2V}
  L_{m\rightarrow n}^{\text{C}} = 2\tanh^{-1}\left(\prod_{n'\in\mathcal{N}(m)\setminus n} \tanh\left(\frac{1}{2} L_{n'\rightarrow m}^{\text{V}}\right) \right).
\end{eqnarray}
The total LLR of $v_n$
\begin{eqnarray}\label{eqn:VLLR}
    L_n = \frac{2y_n}{\sigma^2}+\sum_{m\in\mathcal{M}(n)} L_{m\rightarrow n}^{\text{C}},
\end{eqnarray}
is used to make tentative decoding decision $\hat{u}_{n}=\BSGN(L_n)$, where $\BSGN(a)=0$ if $a\geq 0$ and $\BSGN(a)=1$ otherwise.

The original RBP algorithm \cite{RBP} repeats the following four-step message updating procedure:\newline
\indent \textbf{1)} Compute the C2V messages $\tilde{L}_{m\rightarrow n}^{\text{C}}$ by (\ref{eqn:C2V})
for all $(m,n)$ where $h_{mn}=1$ and the corresponding C2V message residuals (also referred to as C2V residuals for simplicity) by
\begin{eqnarray}\label{eqn:C2VRSD}
  R_{m\rightarrow n}^{\text{C}} = |\tilde{L}_{m\rightarrow n}^{\text{C}}-L_{m\rightarrow n}^{\text{C}} |.
\end{eqnarray}
\indent \textbf{2)} Determine the C2V edge to be updated
\begin{equation}\label{RBP}
  (m^*, n^*)=\arg \max_{(m,n)}R_{m\rightarrow n}^{\text{C}}.
\end{equation}
\indent \textbf{3)} Perform the sole update
\begin{equation}
L_{m^*\rightarrow n^*}^{\text{C}} \leftarrow \tilde{L}_{m^*\rightarrow n^*}^{\text{C}}.
\end{equation}
\indent \textbf{4)} After the updated C2V message is received by $v_{n^*}$, the decoder updates and propagates the V2C messages
$L_{n^*\rightarrow i}^{\text{V}}$, $i \in \mathcal{M}(n^*) \setminus m^*$ based on (\ref{eqn:V2C}).

As only the C2V message $L_{m^*\rightarrow n^*}^{\text{C}}$ is updated and sent, we refer to $\{\tilde{L}_{m\rightarrow n}^{\text{C}}\}$
as the precomputed C2V messages. A decoding iteration is counted after $E$ C2V messages are propagated, where $E$ is the number
of edges on the code graph. The decoder makes tentative codeword check at the end of each iteration and stops when a valid codeword
is found or the maximum iteration number has been reached.

\subsection{Other Scheduling Strategies}\label{subsection_SVNF}
As mentioned before, several improved RBP algorithms have made an effort to avoid updating a small group of
edges repeatedly. In particular, the node-wise RBP \cite{RBP} decoder allows simultaneously updates
of more than one C2V message, the quota-based RBP \cite{SVNF} limits each edge's update times per
iteration and the silent-variable-node-free RBP (SVNF-RBP) method \cite{SVNF} requires that every VN's
intrinsic message should be passed to a connecting CN with a fixed updating order. The dynamic SVNF-RBP (DSVNF-RBP)
algorithm \cite{IDS17a} relaxes the fixed updating order constraint. The residual-decaying-based RBP algorithm
\cite{IDS19a} scales the residual value of a message by a factor which decays with the
number of times the same edge has been updated, thereby reducing the probability of its further update within
an iteration. Among these derivatives of the RBP algorithm, we found that, for many practical LDPC codes,
the SVNF-RBP algorithm not only provides improved decoding performance but is computational efficient.

Besides preventing the greedy updating behavior, many schedules were designed to enhance the decoding efficiency by
detecting unreliable tentative VN decisions as soon as possible. For example, one can locate the VNs which are
likely to have incorrect LLR signs and give them higher updating priority \cite{IDS17a}, \cite{IDS11a}-\cite{IDS19b}.
When a VN is updated, it automatically sends V2C messages to its connecting CNs like Step \textbf{4)} of the RBP algorithm.
The DSVNF-RBP algorithm \cite{IDS17a} first considers those C2V edges connecting to the unsatisfied CNs. As an unsatisfied
CN must link to at least one incorrect VN decision, updating the edges participating in unsatisfied CNs may help reversing
the erroneous decisions.

In \cite{IDS11a}-\cite{IDS16a}, the reliability of $\hat{u}_n$ is judged by checking if it changes sign after an update.
Let $\tilde{L}_n=2y_n/\sigma^2+\sum_{m\in\mathcal{M}(n)} \tilde{L}_{m\rightarrow n}^{\text{C}}$ be the precomputed
LLR of $v_n$. In \cite{IDS11a} and \cite{IDS15a}, a VN's tentative decision $\BSGN(L_n)$ is regarded as unstable if
$\BSGN(L_n)\neq\BSGN(\tilde{L}_n)$ and the unstable VNs are given higher updating priority. In \cite{IDS16a}, a VN's
reliability is judged by checking if the associated tentative bit decisions remain unchanged in three consecutive updates.
Among the unreliable VNs, the one with the largest total VN LLR difference $|\tilde{L}_n -L_n |$ is chosen for update.
In \cite{IDS19b}, the VNs are further classified into four types according to a certain decision reliability metric
so that the most unreliable VN can be updated by using the most reliable local messages on the code graph.

As mentioned in the previous section, a decoding schedule should be VN-centric and focus on selecting an edge which can
help its connected VN to make a better bit decision. We thus opt to have a schedule that prioritizes improving the most
unreliable VN decisions. We adopt an VN-then-edge strategy which determines the targeted VN and from its connecting edges,
select one for C2V message update. How such an approach serves our design goal will become clear in the subsequent discourse.

%% file: Sec3_PDRBP.tex
\section{Conditional Innovation and CIRBP Decoding}\label{section:PD}
\subsection{CI and VN Decisions}
Updating the VNs with unreliable bit decisions to enhance the chance of reversing erroneous decisions
can significantly improve both the convergence speed and the converged error rate. This perhaps is the
rationale behind some related works \cite{IDS11a}-\cite{IDS16a} that prioritize updating the unreliable
VNs (i.e., the decision-changed VNs). The reliability metric used there was derived from the stability
of the VN decisions. Updating an unstable VN with the largest total LLR change may help reversing the bit
decision but not necessarily toward the correct one. Furthermore, this metric tends to ignore unreliable
VNs that have stable decisions but small LLR magnitudes and reduce their chances for improving reliability.
Hence, we need a metric that avoids these shortcomings and, ideally, we would like this metric to be able
to accurately predict the degree of a VN decision's correctness and its chance of being corrected if updated.
In the following paragraphs, we present a metric which possesses similar properties.

Define $\mathcal{O}_Z=\{0,1,\ldots,Z-1\}$ where $Z\in \mathbb{Z}^+$
and $p_{\text{e},n} = \Pr(\hat{u}_n \neq u_n)$ as the bit error probability of the current
decision $\hat{u}_n$.
The codeword error probability would be
\begin{eqnarray}\label{eqn:FER}
    \Pr( \hat{\bm{u}} \neq \bm{u} ) = 1-\prod_{n\in\mathcal{O}_N} \left( 1 - p_{\text{e},n} \right).
\end{eqnarray}
Analogously, we denote by $\tilde{u}_n$ the decision after $v_n$ is updated (i.e., $\tilde{u}_n=\BSGN(\tilde{L}_n)$)
and let $\tilde{p}_{\text{e},n}=\Pr(\tilde{u}_n \neq u_n )$. If only one VN is updated at one time and both $p_{\text{e},n}$
and $\tilde{p}_{\text{e},n}$ were available, to maximally lower the codeword error probability, it is reasonable to select
a VN $v_{n^*}$ which has the best chance of improving its bit error probability for update. That is,
\begin{eqnarray}\label{eqn:BERdown}
  n^*=\arg \max_{n \in \mathcal{O}_N} (p_{\text{e},n} - \tilde{p}_{\text{e},n}).
\end{eqnarray}
Since $p_{\text{e},n}$ and $\tilde{p}_{\text{e},n}$ are not available, we seek for an alternate parameter which can help
infer the quantity $(p_{\text{e},n} - \tilde{p}_{\text{e},n})$. We define the conditional posterior probabilities, $\Pr(u_n=0|L_n)=\exp(L_n)/(1+\exp(L_n))\stackrel{def}{=}p_0(L_n)$ and $\Pr(u_n=1|L_n)=1-p_0(L_n)\stackrel{def}{=}p_1(L_n)$;
both are deterministic function of $L_n$ and their values lie within $[0,1)$. The proposed metric
\begin{eqnarray}\label{eqn:PDdef}
  D_n=| p_0(L_n) - p_0(\tilde{L}_n) | = | p_1(L_n) - p_1(\tilde{L}_n)|,
\end{eqnarray}
measures the new information about $u_n$ we may obtain if the update $L_n \leftarrow \tilde{L}_n$ is carried out.
$0\leq D_n<1$ is thus called the conditional innovation (CI) henceforth.

The usefulness of CI is derived from two interesting properties. For convenience, the messages $\{L_n\}$, $\{\tilde{L}_n\}$
and $\{ D_n \}$ are respectively modeled as random variables $L$, $\tilde{L}$ and $D$. $P_0 = \exp(L)/(1+\exp(L))$,
$P_1 = 1-P_0$, and $\tilde{P}_0$ and $\tilde{P}_1$ are similarly defined for $\tilde{L}$. The first property has to do with
the behavior of the function
\begin{eqnarray}  \nonumber
  \mathcal{J}(\gamma)
  &\triangleq& \frac{\Pr \left( \text{the decision is correct} | D\geq \gamma \right) }{ \Pr \left( \text{the decision is incorrect} | D\geq \gamma \right) }
  \\
  &=& \label{obj_func}
  \frac{\Pr \left( P_0\geq 0.5 | D \geq \gamma \right) }{ \Pr \left( P_0<0.5 | D\geq \gamma \right) },
\end{eqnarray}
where the second equality holds by assuming that the all-zero codeword is transmitted. This assumption is used throughout our analysis
without explicitly mentioned or appeared in related conditional probability expressions.

In \ref{app:A}, we apply the Gaussian approximation (GA) based density evolution (DE) technique \cite{DEol} to show that
\begin{property}
When the BP algorithm is applied to decode an LDPC code in AWGN channels and the C2V messages can be modelled as i.i.d. Gaussian random variables, $\mathcal{J}(\gamma)$, is a decreasing function of the threshold $\gamma$ when the signal-to-noise ratio (SNR) is sufficient large.
\end{property}

\begin{figure}
    \centering
    \subfigure[\label{PD_plot_GA8000} $\mathcal{J}(\gamma)$ obtained by GA-DE with rate=0.5, ($d_v$=4,$d_c$=8) and simulated $\mathcal{J}(\gamma)$ for Gallager (8000,4000) ($d_v$=4,$d_c$=8) code.]
    {\epsfxsize=3.15in
    \epsffile{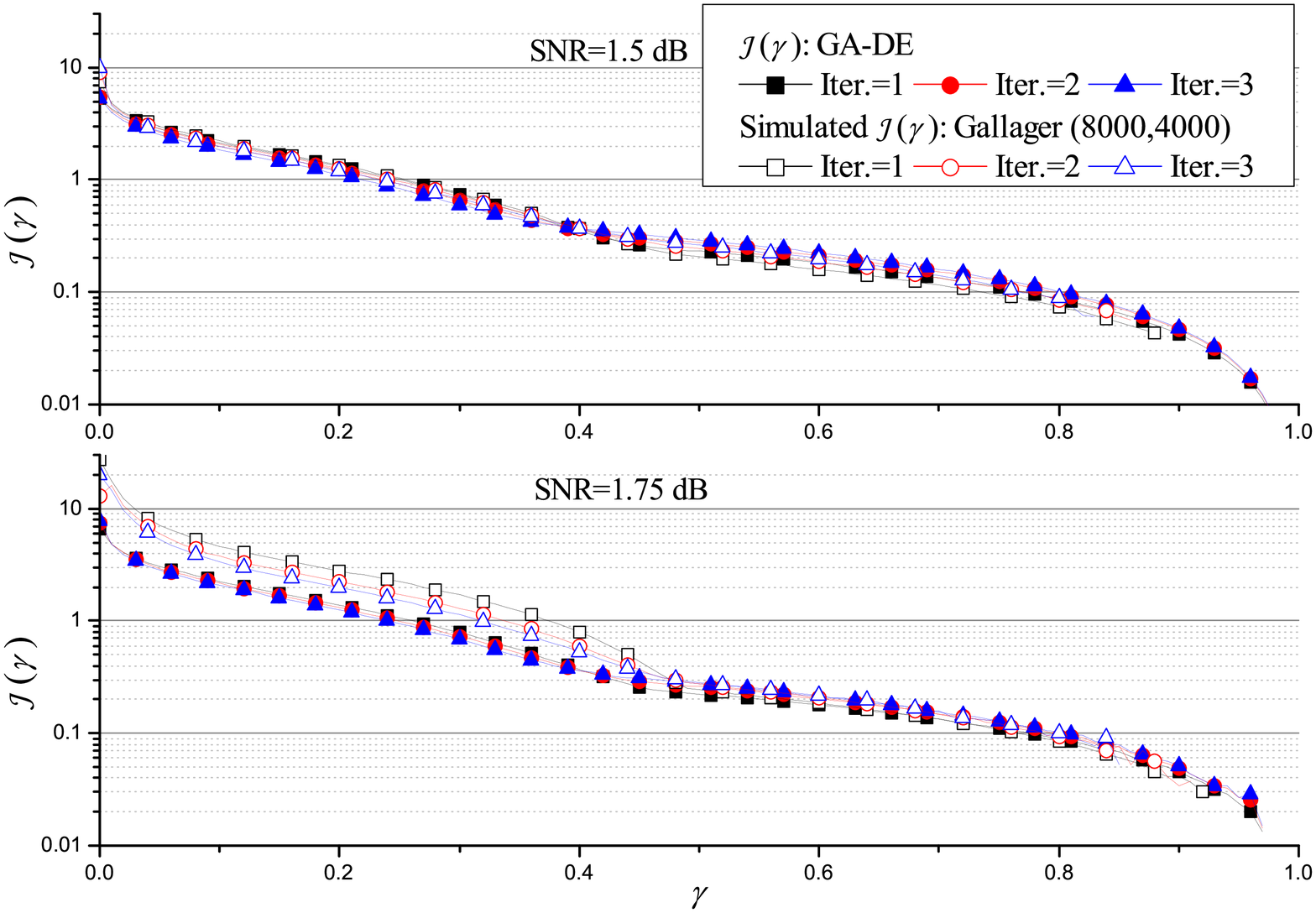}}
    \hspace{0.01\textwidth}
    \subfigure[\label{PD_plot_W1944} Simulated $\mathcal{J}(\gamma)$ for the 802.11 (1944,972) code.]
    {\epsfxsize=3.15in
    \epsffile{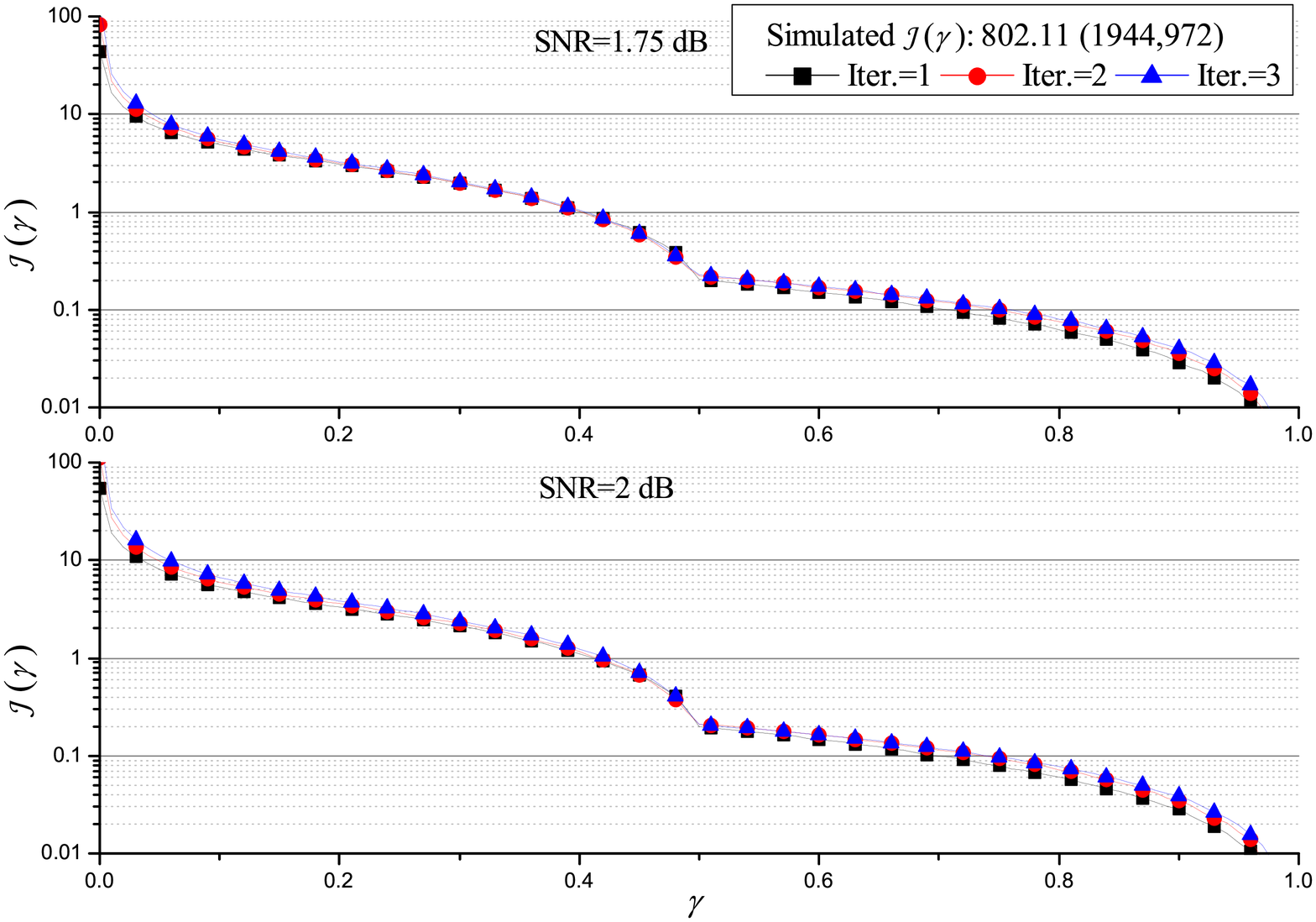}}
    \caption{$\mathcal{J}(\gamma)$ obtained by GA-DE and simulation.}
    \squeezeup
\end{figure}

The i.i.d. C2V messages assumption is the same as that proposed in \cite{DEol} and our proof is semi-analytic in the sense that some parts of
the proof require computer based calculation. The assumption of independent C2V messages \cite{DEol} is valid if the LDPC code of concern is
either cycle-free or the iteration number of interest is smaller than half of the girth of the code so that the VNs do not receive correlated
information.
As an example, we consider a rate-$0.5$ regular code ensemble with CN degree $d_c=8$ and VN degree $d_v=4$. We depict $\mathcal{J}(\gamma)$
of the first three iterations for the flooding schedule in Fig. \ref{PD_plot_GA8000}. For comparison, we also present $\mathcal{J}(\gamma)$
for the $(8000,4000)$ Gallager code with $(d_c,d_v)=(8,4)$ \cite{MacKayWeb} where the corresponding conditional probabilities are obtained
by simulation. In Fig. \ref{PD_plot_W1944}, we show the simulated $\mathcal{J}(\gamma)$ for the 802.11 (1944,972) code. For both cases,
we find that $\mathcal{J}(\gamma)$ is a decreasing function of $\gamma$ and, for a VN whose $D_n$ is sufficiently large, the associated
bit decision is likely to be incorrect. 
We further prove in \ref{app:B} that
\begin{property}
Under the same assumptions of {\it Property 1},
if the current decision is incorrect ($\hat{u}_n \neq u_n$), i.e., for all $P_0 < 1/2$, the function
\begin{eqnarray}\label{app_eq1}
  F(\gamma) \triangleq
  \frac
  {\Pr  ( \tilde{P}_0 \geq P_0 ~|~ D \geq \gamma )}
  {\Pr  ( \tilde{P}_0 < P_0 ~|~ D \geq \gamma )}
\end{eqnarray}
is always larger than $1$, and it is a strictly increasing function of $\gamma$ when $\gamma\in[0,P_0)$ and goes to infinity when $\gamma\in[P_0,1)$.
\end{property}

This property implies that if the bit decision of a VN is incorrect, the larger the associated $D_n$ is,
the greater the probability of making a correct decision after an update becomes, that is,
$\Pr (p_0(\tilde{L}_n)>p_0(L_n))$ increases. These two properties indicate that we should give the VN
with the largest $D_n$ the highest updating priority. This VN has the highest probability of being both
incorrect (before update) and correctable (after update).

\subsection{The CIRBP Decoding Algorithm}

\begin{algorithm}
    \caption{Conditional Innovation Based RBP (CIRBP) Algorithm} \label{PDRBP}
    \label{PDRBP_algo}
    \begin{algorithmic}[1]
    \State Initialize all $L_{m\rightarrow n}^{\text{C}}=0$ and all $L_n=L_{n\rightarrow m}^{\text{V}}=2y_n/\sigma^2$
    \squeezeupAlgo
    \State Generate all $\tilde{L}_{m\rightarrow n}^{\text{C}}$ by (\ref{eqn:C2V}) and compute all $R_{m\rightarrow n}^{\text{C}}$
    \squeezeupAlgo
    \State Compute all $\tilde{L}_n$ and $D_n$
    \squeezeupAlgo
    \State Find $n^*=\arg\max_j \{D_j~|~j\in\mathcal{O}_N\}$
    \label{ALG:PDRBP_findVN}
    \squeezeupAlgo
    \If{$D_{n^*} < \gamma$}
        \squeezeupAlgo
        \State Find $(m^*,n^*)=\arg\max_{(i,j)}\{R_{i\rightarrow j}^{\text{C}}~|~h_{ij}=1\}$ 
         and go to line \ref{ALG:PDRBP_Prop}
    \squeezeupAlgo
    \EndIf
    \squeezeupAlgo
    \State Find $m^*=\arg\max_i \{R_{i\rightarrow n^*}^{\text{C}}~|~i\in\mathcal{M}(n^*)\}$
    \label{ALG:PDRBP_findCN}
    \squeezeupAlgo
    \State Let $L_{m^*\rightarrow n^*}^{\text{C}}\leftarrow \tilde{L}_{m^*\rightarrow n^*}^{\text{C}}$. Propagate $L_{m^*\rightarrow n^*}^{\text{C}}$, let $R_{m^*\rightarrow n^*}^{\text{C}}=0$, and update $L_{n^*}$
    \label{ALG:PDRBP_Prop}
    \squeezeupAlgo
    \For{every $i\in\mathcal{M}(n^*)\backslash m^*$}
        \squeezeupAlgo
        \State Generate and propagate $L_{n^*\rightarrow i}^{\text{V}}$
        \squeezeupAlgo
        \State Compute $\tilde{L}_{i\rightarrow j}^{\text{C}}$, $R_{i\rightarrow j}^{\text{C}}$, $\tilde{L}_j$ and $D_j ~\forall j\in\mathcal{N}(i)\backslash n^*$
    \squeezeupAlgo
    \EndFor
    \squeezeupAlgo
    \State Go to line \ref{ALG:PDRBP_findVN} if \emph{Stopping Condition} is not satisfied
    \end{algorithmic}
\end{algorithm}

Based on the above discussion, we propose the CI based RBP (CIRBP) algorithm as shown in {\bf Algorithm \ref{PDRBP}}.
The VN with the largest CI can be selected as the candidate VN for update as it is the VN which is most likely to yield
an erroneous decision if $D_{n^*}\geq \gamma$ and it is also the most correctable if updated; otherwise, identifying the
incorrect decision(s) becomes difficult and the C2V update would then follow the original RBP algorithm.
Such threshold-based judgement is based on our observation in Figs. \ref{PD_plot_GA8000} and \ref{PD_plot_W1944} that the probability
that a VN decision is wrong is a monotonic decreasing function of $\gamma$. For the selected VN, denoted by $v_{n^*}$ henceforth,
the associated incoming C2V message $L^\text{C}_{m^*\rightarrow n^*}$, which has the maximum residual, is updated (lines 8--9). Using
this C2V message, $v_{n^*}$ then sends new V2C messages to $c_i$, $i \in \mathcal{M}(n^*)\setminus m^*$ and the associated messages $\tilde{L}_{i\rightarrow j}^{\text{C}}$, $R_{i\rightarrow j}^{\text{C}}$, $\tilde{L}_j$ and $D_j~\forall j\in\mathcal{N}(i)\backslash n^*$, will be calculated (lines 10--13).

As $\mathcal{J}(\gamma)$ depends on the code structure, the iteration number, SNR and the decoding schedule used and is not admitted
in a closed-form expression. Its monotonicity property can only be proved semi-analytically. For practical concerns, we use a fixed
$\gamma$ and find that a properly chosen $\gamma$ suffices to give outstanding performance. The chosen $\gamma$ cannot be too small
for then CI is no longer a reliable indicator in identifying the incorrect yet correctable bit decision. But if $\gamma$ is too large,
the probability $\Pr(D_{n^*}\geq \gamma)$ becomes very small and our CIRBP decoder will rely on the conventional LLR residual most of
the time and gives diminishing gain against the original RBP decoder.

%% file: Sec4_LMDRBP.tex
\section{Latest-Message-Driven Schedule and LMDRBP Algorithms}\label{section:LMD}
\subsection{LMD Scheduling Strategy}\label{subsection_LMD}

Most IDS strategies focus on using some message reliability metric to select the C2V messages to be propagated.
On the other hand, the update criteria presented in the previous section and in \cite{IDS17a}, \cite{IDS19b},
are implicitly designed to select a VN such that the selected one can make a better bit decision. Both
approaches eventually improve the reliability of the V2C messages which the target VN is going to deliver and
the resulting decoders do yield performance better than that of the standard BP decoder with the same iteration
or edge update number. It is reasonable to conjecture that not only the V2C messages emitted from the latest updated VN ($v_{n^*}$)
but also the subsequent C2V messages forwarded by the connecting CNs become more trustworthy. This conjecture
suggests that the decoding schedule prioritize using the messages originated from those nodes which are just
updated and possess the newest information. An extra benefit of considering only newly updated nodes and
messages is the reduction of the search range for finding a suitable C2V message or VN for the next update.

Based on this idea and following the VN-centric guideline, we propose the latest-message-driven (LMD) RBP (LMDRBP)
algorithm as described in \textbf{Algorithm} \ref{LMDRBP_algo}. In this algorithm we compare the C2V residuals of the
latest renewed C2V messages, i.e., the messages forwarded by those CNs which just received new V2C messages from the
latest-updated VN, and select the VN $v_{n^*}$ associated with the maximum C2V residual as the next update target.
For the selected VN, we compare all its connected C2V messages---both new and old---and accept only the one with the
largest residual (lines 9--11). By doing so, we reduce the VN search range to the nearest neighboring VNs of the latest
updated VN but not the C2V message search range of the targeted VN and avoid favoring a certain group of edges. The
total LLR of $v_{n^*}$ and the associated V2C messages are updated, and then the CNs linking to $v_{n^*}$ precompute
their C2V messages and residuals to complete an update procedure (lines 4--8). This procedure repeats until the
stopping condition is satisfied. The numerical results presented in the next section show that \textbf{Algorithm}
\ref{LMDRBP_algo} outperforms most existing RBP algorithms, indicating the important fact that its search range
reduction effort not only significantly eases the search load but also filters many improper candidate nodes/edges
from its search list and thus lowers the probability of making a wrong update selection.
\begin{algorithm}
    \caption{Latest-Message-Driven RBP (LMDRBP) Algorithm}
    \label{LMDRBP_algo}
    \begin{algorithmic}[1]
    \State Initialize all $L_{m\rightarrow n}^{\text{C}}=0$ and all $L_{n\rightarrow m}^{\text{V}}=2y_n/\sigma^2$
    \squeezeupAlgo
    \State Generate all $\tilde{L}_{m\rightarrow n}^{\text{C}}$ by (\ref{eqn:C2V}) and compute all $R_{m\rightarrow n}^{\text{C}}$
    \squeezeupAlgo
    \State Find $(m^*,n^*)=\arg\max_{(i,j)}\{R_{i\rightarrow j}^{\text{C}}~|~ h_{mn}=1\}$
    \squeezeupAlgo
    \State Let $L_{m^*\rightarrow n^*}^{\text{C}}\leftarrow \tilde{L}_{m^*\rightarrow n^*}^{\text{C}}$. Propagate $L_{m^*\rightarrow n^*}^{\text{C}}$, let $R_{m^*\rightarrow n^*}^{\text{C}}=0$, and update $L_{n^*}$
    \label{ALG:LMDRBP_Prop}
    \squeezeupAlgo
    \For{every $i\in\mathcal{M}(n^*)\backslash m^*$}
        \squeezeupAlgo
        \State Generate and propagate $L_{n^*\rightarrow i}^{\text{V}}$
        \squeezeupAlgo
        \State Compute $\tilde{L}_{i\rightarrow j}^{\text{C}}$ and update $R_{i\rightarrow j}^{\text{C}}~\forall j\in\mathcal{N}(i)\backslash n^*$
    \squeezeupAlgo
    \EndFor
    \squeezeupAlgo
    \State Find $(m',n')=\arg\max_{(i,j)}\{R_{i\rightarrow j}^{\text{C}}~|~ i\in\mathcal{M}(n^*)\backslash m^* , j\in \mathcal{N}(i)\backslash n^*\}$
    \squeezeupAlgo
    \State Find $\hat{m}=\arg\max_{i}\{R_{i\rightarrow n'}^{\text{C}}~|~i\in\mathcal{M}(n')\}$ and let $m'\leftarrow \hat{m}$
    \squeezeupAlgo
    \State Let $(m^*,n^*)\leftarrow (m',n')$
    \squeezeupAlgo
    \State Go to line \ref{ALG:LMDRBP_Prop} if \emph{Stopping Condition} is not satisfied
    \end{algorithmic}
\end{algorithm}

In the LMD schedule, both finding the target VN and deciding which C2V message it should receive require real-number comparisons.
To reduce the comparison effort, we bypass line $10$ of \textbf{Algorithm \ref{LMDRBP_algo}} and send the C2V message corresponding
to the maximum residual found in line $9$. This modified version is referred to as the simplified LMDRBP (sLMDRBP) algorithm.

\subsection{LMD-based CIRBP Algorithm}\label{subsection_LMD-PD}
If the information carried by the latest updated C2V messages is more reliable, the related precomputed VN total LLRs ($\tilde{L}_n$'s)
and the CI values can also be more trustworthy after incorporating these newest messages. Combining the concepts of the LMD schedule
and the CIRBP decoder can then improve the accuracy of the VN reliability judgement. Since the CIRBP decoder selects the target VN by
comparing VNs' CI values, we modify the LMD based schedule by letting the updated VN be decided by the last-updated CI values instead of
the C2V residuals. With the modified schedule, we have LMD-based CIRBP (LMD-CIRBP) decoding algorithm described in \textbf{Algorithm \ref{LMD-PDRBP_algo}}.

To determine the initial updated VN and edge, we simply select the VN with the global maximum CI be the initial targeted VN (line 4).
The initial chosen edge will be the one which has the maximum residual among all candidate C2V messages to be sent to the targeted VN (line 5).
Let $L_{m^*\rightarrow n^*}^{\text{C}}$ be the selected C2V message and $c_{m^*}$ and $v_{n^*}$ respectively be the corresponding CN and VN.
The V2C messages from $v_{n^*}$ (i.e., $L_{n^*\rightarrow i}^{\text{V}}$)
would be updated, and then all associated precomputed messages, C2V residuals, and CIs will also be renewed (lines 7--10).
For all VNs in the set $\mathcal{U}(m^*,n^*)\triangleq \{\tilde{n} | \tilde{n} \in \mathcal{N}(\tilde{m})\setminus n^*, ~\tilde{m}\in \mathcal{M}(n^*)\setminus m^*\}$,
the one with the maximum CI is chosen as the next update target which accepts the C2V message from one of its connecting edges
with the maximum C2V residual (lines 11--12). The above procedure will be repeated until the stopping condition is met.
The LMD-CIRBP algorithm enjoys the advantages of both CIRBP and LMDRBP decoders--it not only has better chance to locate the
VNs which indeed need to be updated but requires much less search complexity since only those CI values for the VNs in
$\mathcal{U}(m^*,n^*)$ need to be compared.
\begin{algorithm}
    \caption{LMD-Based CIRBP (LMD-CIRBP) Algorithm} \label{LMD-PDRBP}
    \label{LMD-PDRBP_algo}
    \begin{algorithmic}[1]
    \State Initialize all $L_{m\rightarrow n}^{\text{C}}=0$ and all $L_{n\rightarrow m}^{\text{V}}=2y_n/\sigma^2$
    \squeezeupAlgo
    \State Generate all $\tilde{L}_{m\rightarrow n}^{\text{C}}$ by (\ref{eqn:C2V}) and compute all $R_{m\rightarrow n}^{\text{C}}$
    \squeezeupAlgo
    \State Compute all $\tilde{L}_n$ and $D_n$
    \squeezeupAlgo
    \State Find $n^*=\arg\max_j \{D_j~|~j\in\mathcal{O}_N\}$
    \squeezeupAlgo
    \State Find $m^*=\arg\max_i \{R_{i\rightarrow n^*}^{\text{C}}~|~i\in\mathcal{M}(n^*)\}$
    \squeezeupAlgo
    \State Let $L_{m^*\rightarrow n^*}^{\text{C}}\leftarrow \tilde{L}_{m^*\rightarrow n^*}^{\text{C}}$, propagate $L_{m^*\rightarrow n^*}^{\text{C}}$, let $R_{m^*\rightarrow n^*}^{\text{C}}=0$, and update $L_{n^*}$
    \label{ALG:LMD-PDRBP_Prop}
    \squeezeupAlgo
    \For{every $i\in\mathcal{M}(n^*)\backslash m^*$}
        \squeezeupAlgo
        \State Generate and propagate $L_{n^*\rightarrow i}^{\text{V}}$
        \squeezeupAlgo
        \State Compute $\tilde{L}_{i\rightarrow j}^{\text{C}}$, $R_{i\rightarrow j}^{\text{C}}$, $\tilde{L}_j$ and $D_j~\forall j\in\mathcal{N}(i)\backslash n^*$
    \squeezeupAlgo
    \EndFor
    \squeezeupAlgo
    \State Find $n'=\arg\max_{j}\{D_j~|~j\in\mathcal{U}(m^*,n^*)\}$.
    \squeezeupAlgo
    \State Let $n^*\leftarrow n'$ and go to line \ref{ALG:LMD-PDRBP_Prop} if \emph{Stopping Condition} is not satisfied
    \end{algorithmic}
\end{algorithm}

%% file: Sec5_Result.tex
\section{Numerical Results and Complexity Analysis}\label{section:simulation}
In this section, we compare the frame error rate (FER) performance and computational complexity of the proposed
and some known RBP decoders. The simulation setup is the same as what was described in Sec. \ref{subsection_BP},
i.e., an LDPC coded data stream is BPSK-modulated and transmitted over an AWGN channel with two-sided power spectral
density $N_0/2=\sigma^2$. Three LDPC codes are considered: the $(1944,972)$ rate-$1/2$ LDPC code of the IEEE 802.11
standard (WiFi) \cite{WiFi}, and the $(1848,616)$ rate-$1/3$ and $(500,100)$ rate-$1/5$ LDPC codes used in the 5G NR
specification \cite{NR}. We denote these codes by W-$1944$, N-$1848$ and N-$500$, respectively. According to 5G NR
specification, N-$1848$ is obtained by puncturing the first 56 VNs of a length-$1904$ mother code generated based on
Base Graph 1 (BG1) with lifting size 28 while N-$500$ is obtained by puncturing the first 20 VNs of a length-$520$
mother code derived from Base Graph 2 (BG2) with lifting size 10. As mentioned in Section \ref{section:review}, an
iteration is defined as $E$ C2V message propagations, and $I_{\text{max}}$ denotes the maximum allowed iteration number.

Both the precomputed and actual propagated C2V messages are calculated by (\ref{eqn:C2V}). If we use the min-sum
approximation \cite{MSA} instead of (\ref{eqn:C2V}) for the C2V message precomputations \cite{RBP}, \cite{SVNF},
the computation load decreases possibly at the expense of performance degradation.

\subsection{FER Performance}\label{subsection_FER}

\begin{figure}
    \centering
    \subfigure[\label{W1944_CIRBP_thrs_new} W-$1944$ code.]
    {\epsfxsize=3.15in
    \epsffile{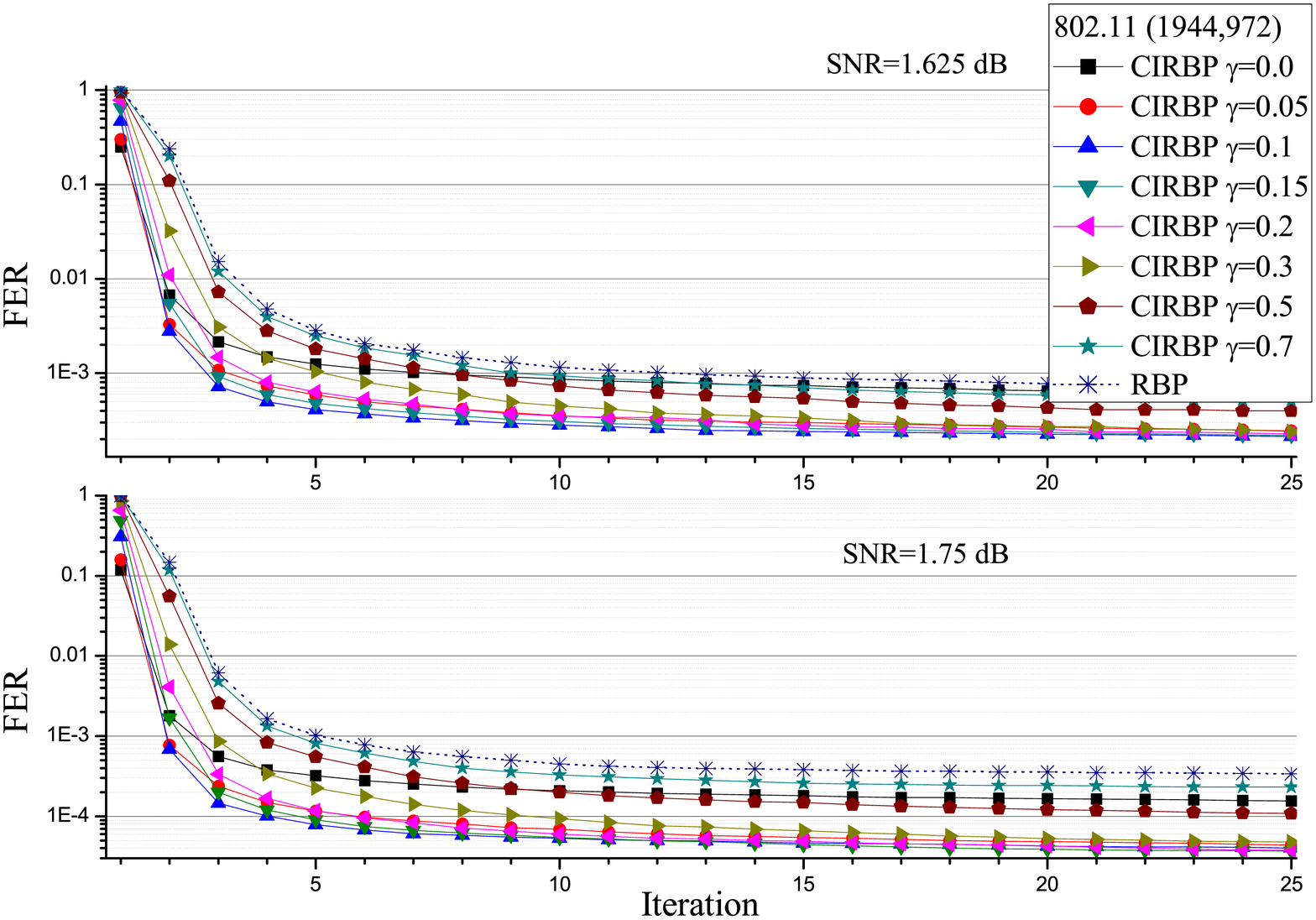}}
    \hspace{0.01\textwidth}
    \subfigure[\label{NR500_CIRBP_thrs_new} N-$500$ code.]
    {\epsfxsize=3.17in
    \epsffile{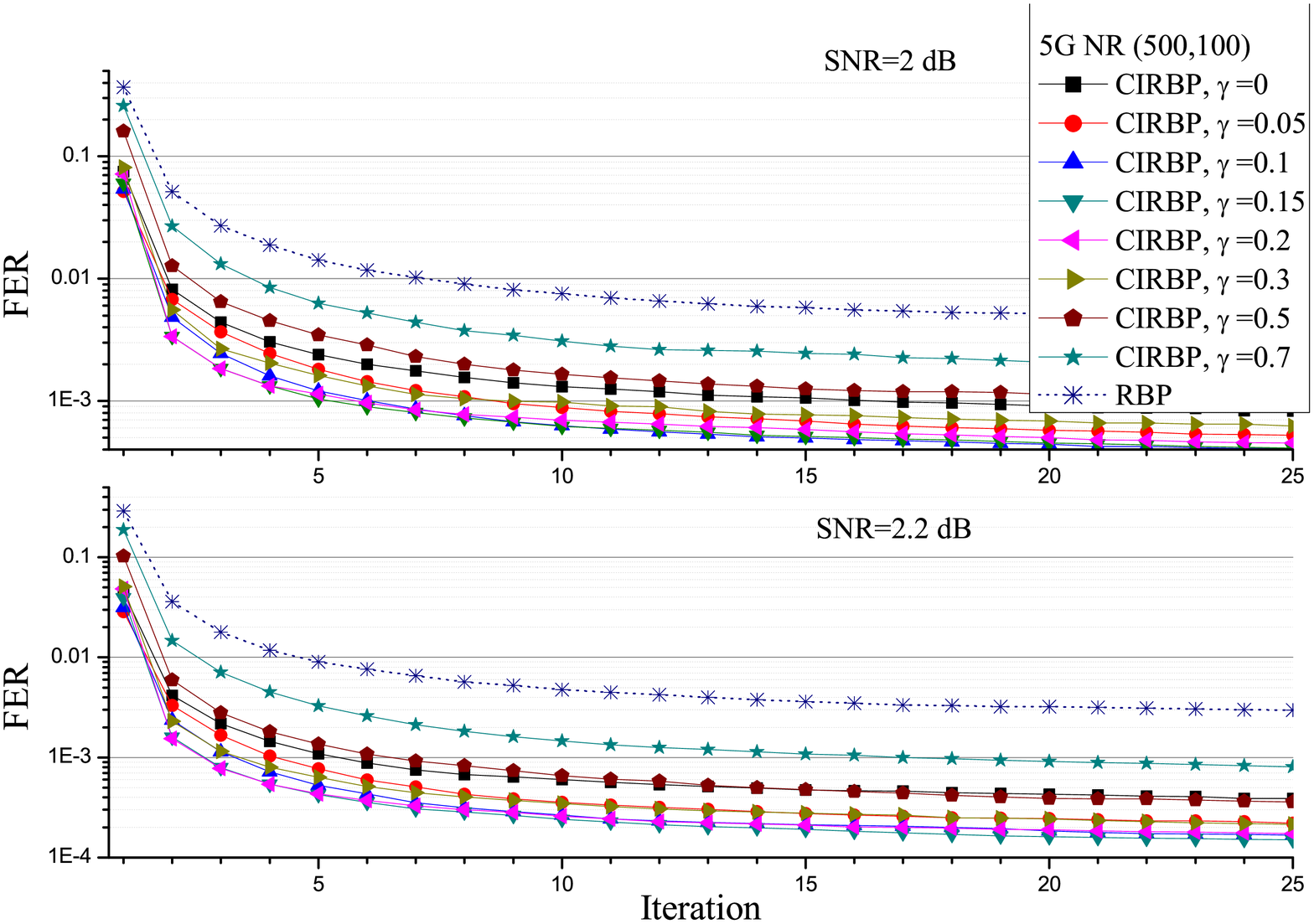}}

    \caption{\label{Fig:CIRBP_gammas}FER convergence behaviors of CIRBP algorithms with different
    $\gamma$ in decoding W-$1944$ and N-$500$ codes}
\end{figure}
In Figs. \ref{W1944_CIRBP_thrs_new} and \ref{NR500_CIRBP_thrs_new}, we show the effect of the CI thresholds ($\gamma$)
on the CIRBP algorithm's performance in decoding the W-$1944$ and N-$500$ codes at different SNRs ($E_b/N_0$). Those
curves indicate that when $\gamma\leq 0.2$, the threshold provides an early-stage and converged performance tradeoff:
$\gamma=0$ or $0.05$ gives the best $1$st-iteration FER performance but $\gamma=0.1$, $0.15$, or $0.2$ results in
better converged FER performance. Although not shown here, our simulations confirm that the N-1848 code renders similar
behaviors. To avoid presenting too many curves in one figure, we only present the CIRBP decoder performance using
$\gamma = 0.1$ and $0.15$ for the remaining figures.

\begin{figure}
    \centering
    \subfigure[\label{W_ite_3} FER and BER performance, $I_{\text{max}}=3$]
    {\epsfxsize=3.15in
    \epsffile{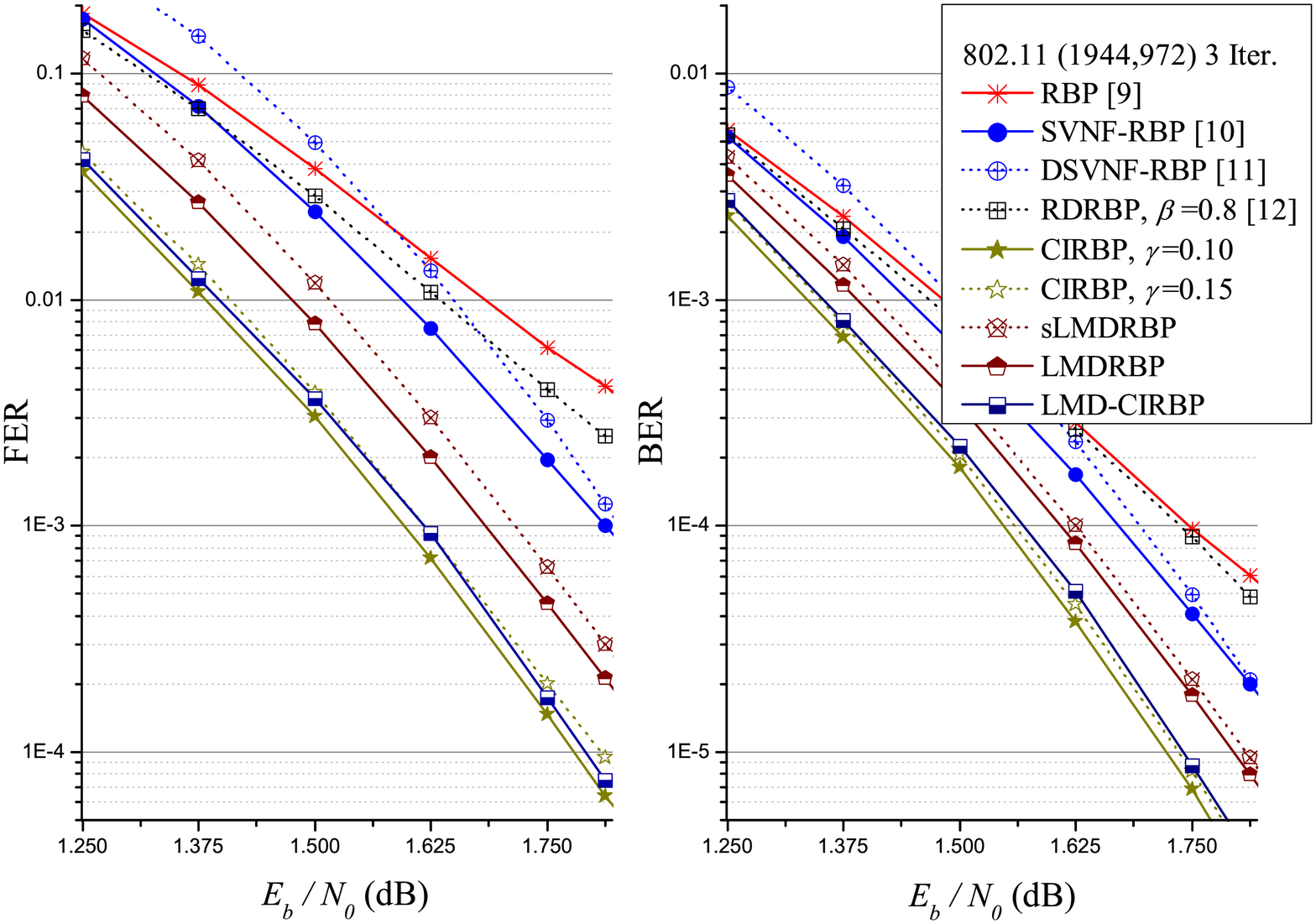}}
    \hspace{0.01\textwidth}
    \subfigure[\label{W_SNR_175} FER and BER convergence behaviors, SNR = $1.75$ dB]
    {\epsfxsize=3.15in
    \epsffile{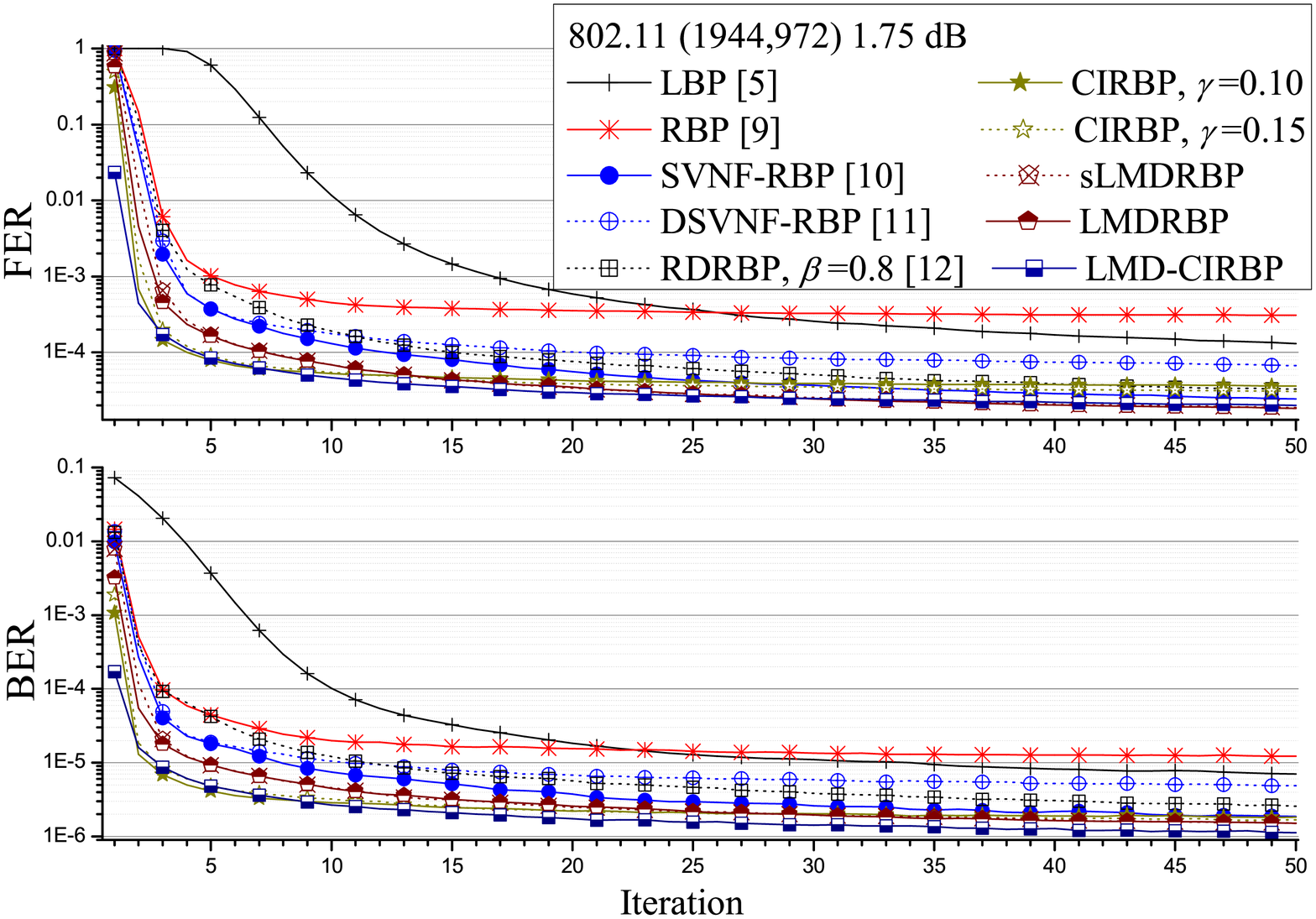}}

    \caption{\label{Fig:W1944}FER and BER performance and convergence behaviors for various IDS-based decoding algorithms; W-$1944$ code.}
    \squeezeup
\end{figure}

Fig. \ref{W_ite_3} plots the error-rate performance of various IDS algorithms in decoding W-$1944$ code
at $I_{\text{max}}=3$. At FER $\approx 10^{-3}$, the CIRBP and LMD-CIRBP algorithms have about $0.2$ dB gain
with respect to the SVNF-RBP and DSVNF-RBP algorithms. The LMDRBP algorithm also outperforms the SVNF-RBP
and DSVNF-RBP algorithms at the same FER. In Fig. \ref{W_SNR_175} we show the FER and BER convergence behaviors
of these algorithms in decoding the same code at SNR $=1.75$ dB. These figures indicate that our algorithms
outperform the SVNF-RBP/DSVNF-RBP (RDRBP) algorithms for $I_{\text{max}}< 30$ ($I_{\text{max}}< 40$).
In addition, the LMDRBP, sLMDRBP and LMD-CIRBP decoders yield better converged ($I_{\text{max}}=50$) FER
performance than that of the SVNF-RBP and RDRBP decoders. Among these decoders, the LMD-CIRBP algorithm gives
by far the best performance at the first iteration.

For the LDPC codes used in IEEE 802.11 systems, the degrees of all VNs are at least two and messages
can be exchanged through every VN. However, there are several degree-1 VNs in the 5G NR codes (N-$500$ and N-$1848$)
and to decode these codes with the LMDRBP and LMD-CIRBP algorithms we have to make some modifications.
More specifically, after the C2V message $L_{m^*\rightarrow n^*}^{\text{C}}$ was sent, the next updated
VN is selected from the VNs which link to $c_i$ for all $i\in\mathcal{M}(n^*)\backslash m^*$.
If $v_{n^*}$ is a degree-1 node,
we allow the decoder to search for the next updated VN from the set $\mathcal{N}(m^*)\backslash n^*$ and
line 9 of \textbf{Algorithm \ref{LMDRBP_algo}} is replaced by ``Find $(m',n')=\arg\max_{(i,j)}\{R_{i\rightarrow j}^{\text{C}}~|~ i\in
\mathcal{M}(n^*) , j\in \mathcal{N}(i)\backslash n^*\}$" while line 11 of \textbf{Algorithm \ref{LMD-PDRBP_algo}}
is to be modified as ``Find $n'=\arg\max_{j}\{D_j~|~j\in\mathcal{N}(m^*)\backslash n^*\}$".

\begin{figure}
    \centering
    \subfigure[\label{NR500_ite_3} FER and BER performance, $I_{\text{max}}=3$]
    {\epsfxsize=3.15in
    \epsffile{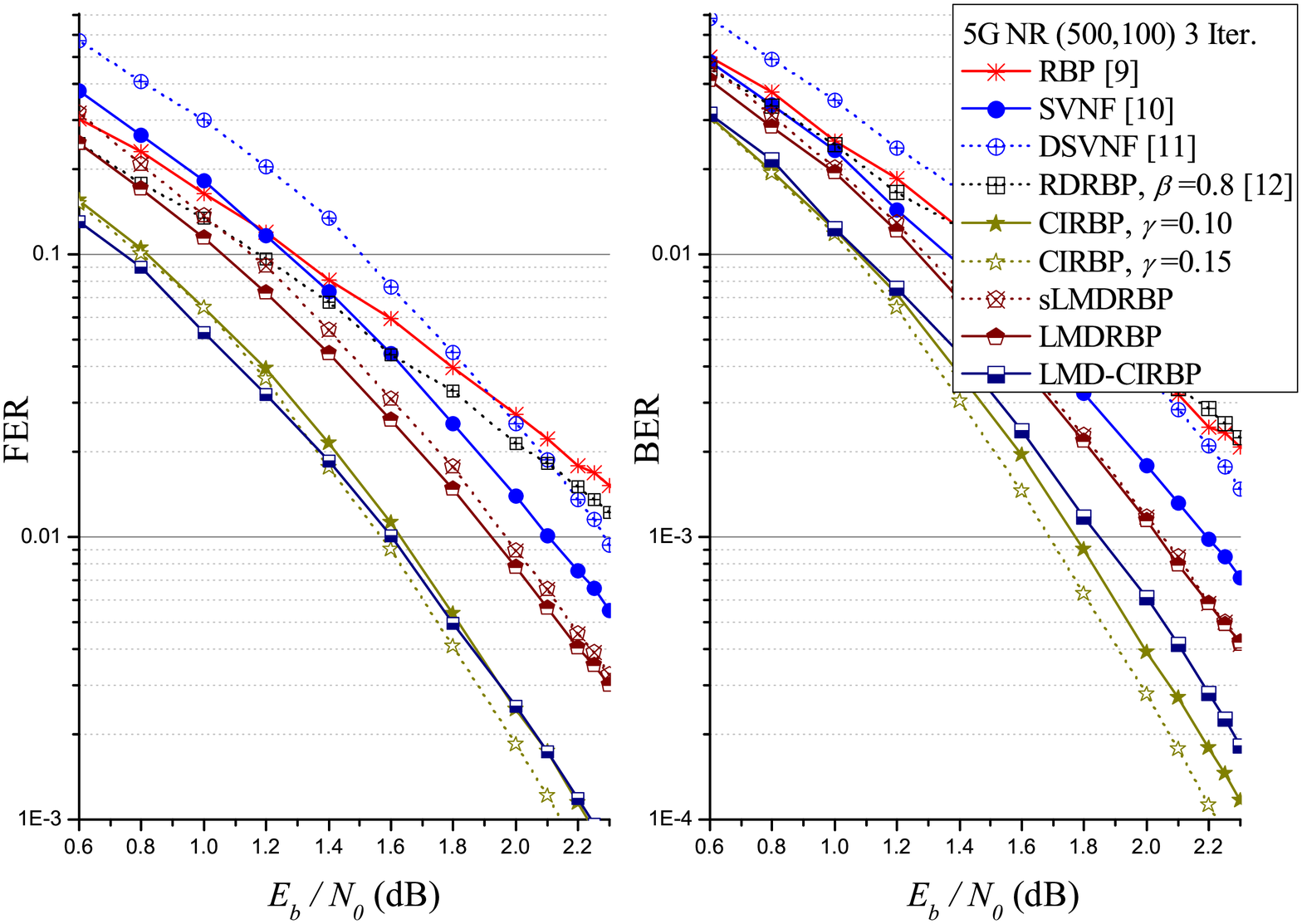}}
    \hspace{0.01\textwidth}
    \subfigure[\label{NR500_SNR_220} FER and BER convergence behaviors, SNR = $2.2$ dB]
    {\epsfxsize=3.15in
    \epsffile{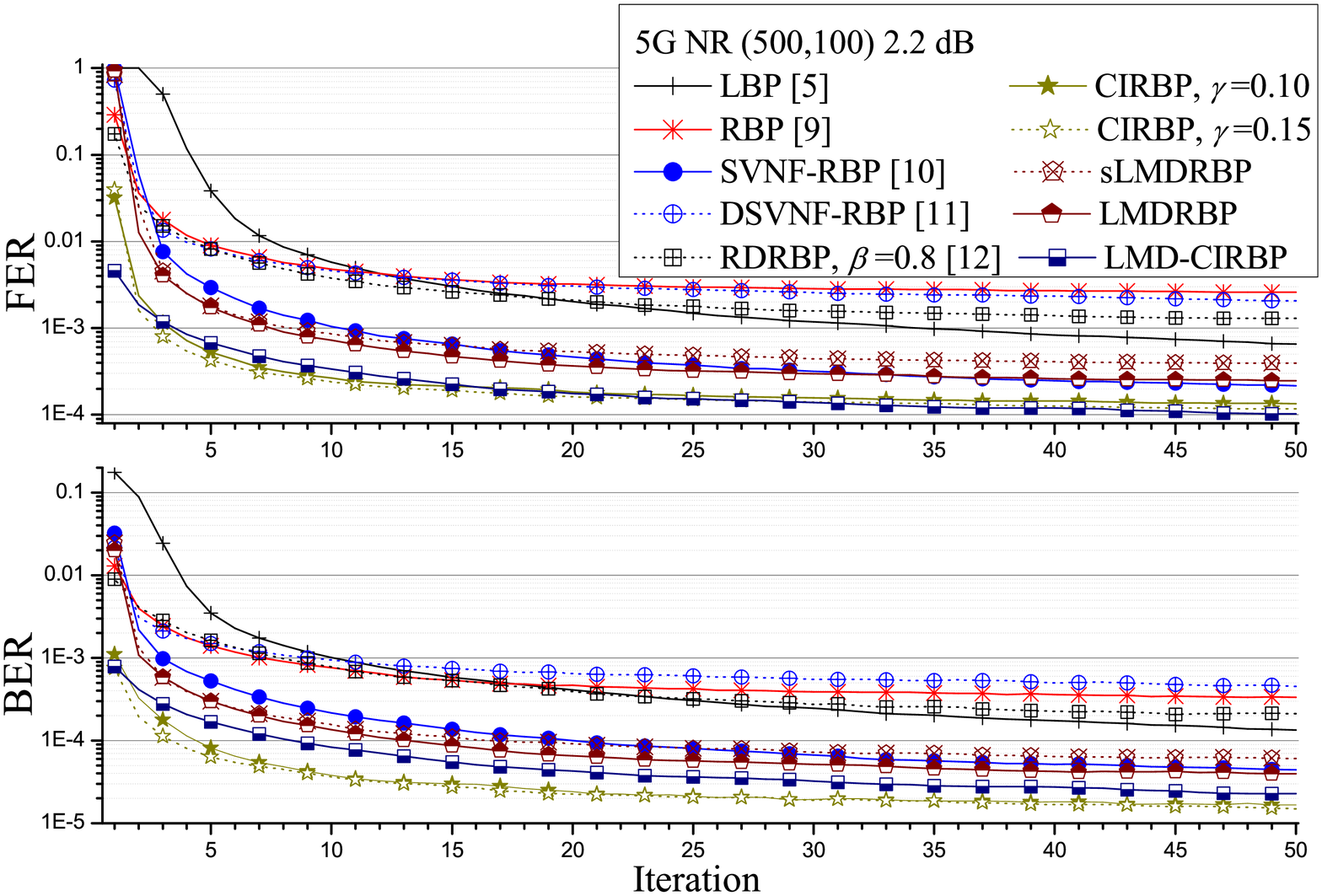}}

    \caption{\label{Fig:N500}FER and BER performance and convergence behaviors for various IDS-based decoding algorithms; N-$500$ code.}
    \squeezeup
\end{figure}

\begin{figure}
    \centering
    \subfigure[\label{NR_ite_3} FER and BER performance, $I_{\text{max}}=3$]
    {\epsfxsize=3.15in
    \epsffile{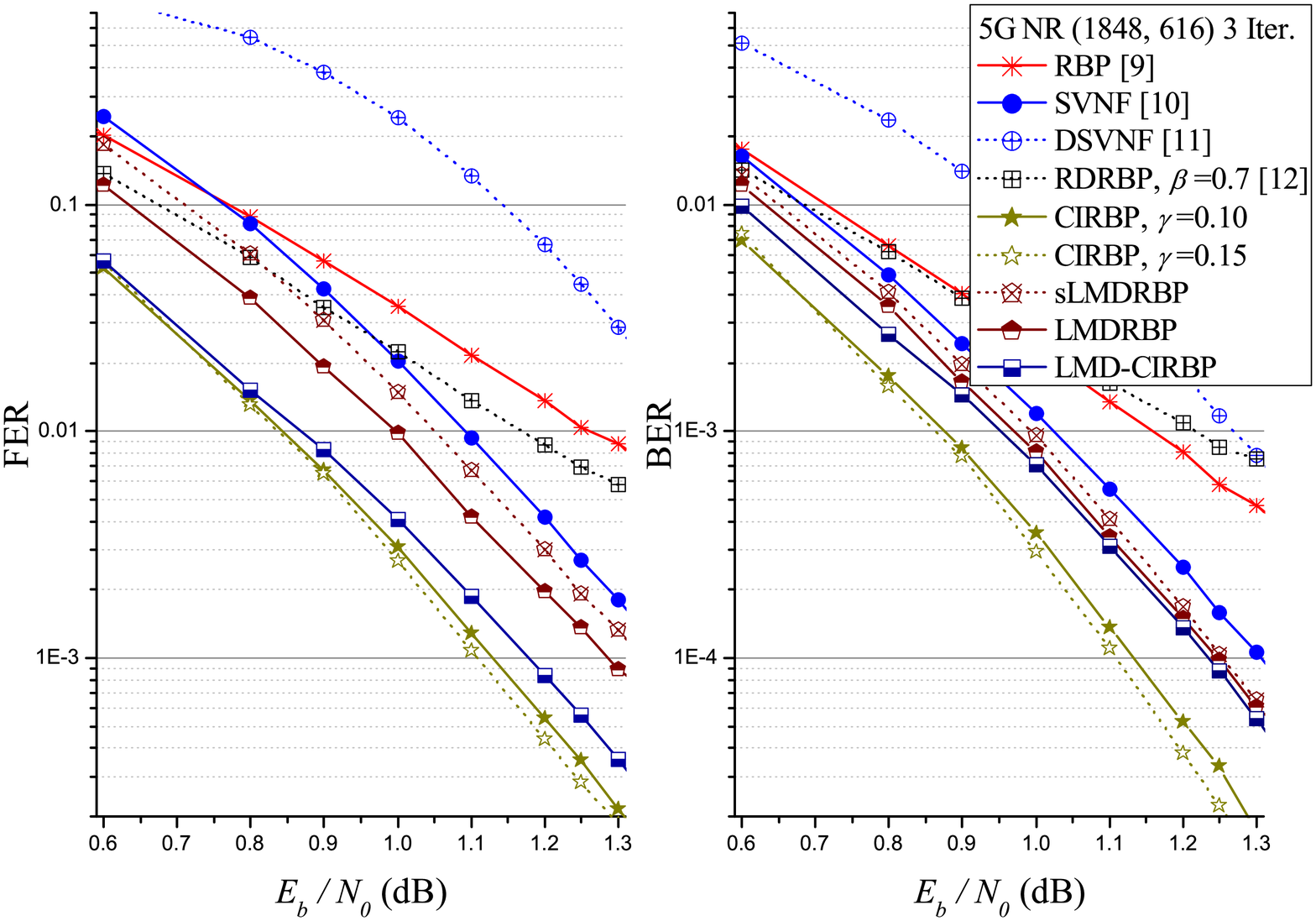}}
    \hspace{0.01\textwidth}
    \subfigure[\label{NR_SNR_130} FER and BER convergence behaviors, SNR = $1.3$ dB]
    {\epsfxsize=3.15in
    \epsffile{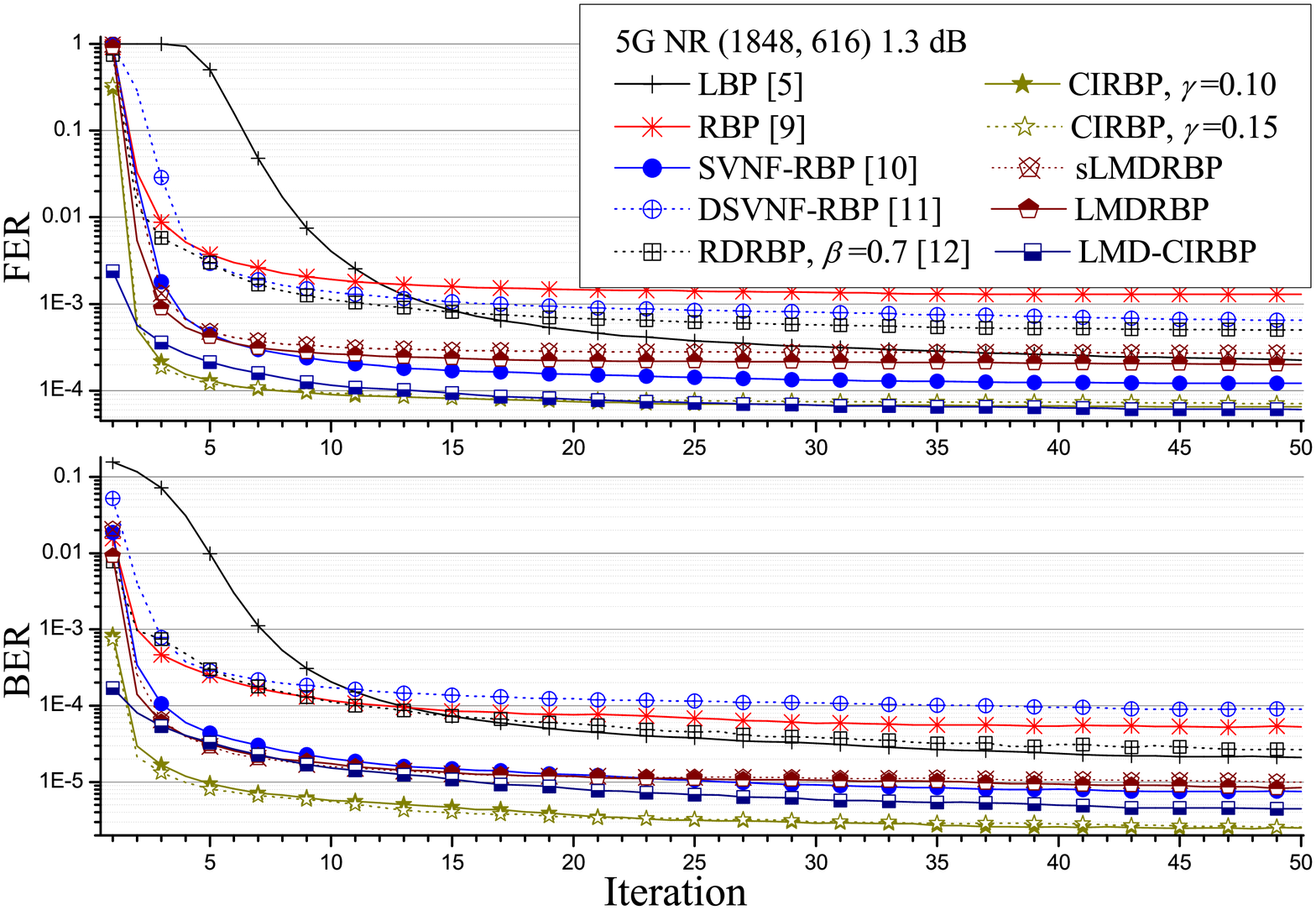}}

    \caption{\label{Fig:N1848}FER and BER performance and convergence behaviors for various IDS-based decoding algorithms; N-$1848$ code.}
    \squeezeup
\end{figure}

Shown in Fig. \ref{NR500_ite_3} is the error-rate performance of various IDS algorithms in decoding the
N-$500$ code with $I_{\text{max}}=3$.  We find that the LMD-CIRBP algorithm outperforms the SVNF-RBP one
by $0.5$ and $0.35$ dB at FER $\approx 10^{-2}$ and BER $\approx 10^{-3}$, respectively. The CIRBP algorithm
provides $0.4$--$0.5$ dB gain in comparison with the SVNF-RBP one. The LMDRBP decoders also give performance
better than that of the SVNF-RBP and DSVNF-RBP decoders.
These decoders' corresponding convergence trends at SNR$=2.2$ dB are shown in Fig. \ref{NR500_SNR_220},
which confirm that our CIRBP and LMD-CIRBP algorithms outperform existing algorithms for all iterations,
and the LMDRBP and sLMDRBP decoders also outperform existing decoders except for the RBP and RDRBP decoders
at the 1st iteration.
It is wroth mentioning that the LMD-CIRBP algorithm's first-iteration performance,
FER $\approx 5\times 10^{-3}$, is quite impressive.

In Figs. \ref{NR_ite_3} and \ref{NR_SNR_130}, we depict the performance of various IDS algorithms with $I_{\text{max}}=3$
and their convergence behaviors at SNR $=1.3$ dB in decoding the N-$1848$ code. We see that, for $I_{\text{max}}=3$
and at FER $\approx 10^{-2}$ or BER $\approx 10^{-4}$, the CIRBP algorithm yields $0.2$ dB gain against the SVNF-RBP
algorithm, and the LMD-CIRBP decoder achieves the same FER gain but has less than $0.1$ dB gain at the same BER.
The LMDRBP decoders still yield performance better than that of the SVNF-RBP algorithm. Fig. \ref{NR_SNR_130} shows
that the CIRBP and LMD-CIRBP algorithms outperform the SVNF-RBP and RDRBP algorithms when $I_{\text{max}}\leq 50$.
With reduced search range, the LMD-CIRBP algorithm still give outstanding first-iteration and converged FER performance.
Although the FER/BER vs. SNR curves are presented for $I_{\text{max}}=3$ only, Figs. 3(b), 4(b) and 5(b) indicate that,
at selected SNRs, the CI-based decoders are better than other IDS-based algorithms for almost all $I_{\text{max}}$ of
interest.

\subsection{Complexity Summary}\label{subsection_compl}

We summarize the decoding complexity of the proposed algorithms and the original RBP, SVNF-RBP, and DSVNF-RBP
algorithms in Table \ref{Complexity Table} in terms of the numbers of required C2V precomputations, CI computations,
and real-number comparisons per update.
In Table \ref{Complexity Table}, a ``C2V pre-update" includes precomputations of C2V messages and residuals,
and a ``CI update" includes computing $\tilde{L}_n$, table look-up of $p_0(\cdot)$, and the evaluation of (\ref{eqn:PDdef})
with a total of three real-number subtractions/additions
involved: two for updating $\tilde{L}_n$ and one for computing the CI.
The ``C2V residual and CI comparisons" counts the real-number comparisons needed for finding the maximum residual and CI value.
$\bar{d}_v$ and $\bar{d}_c$ in Table \ref{Complexity Table} respectively denote average VN and CN degrees,
and ($\bar{d}_v$, $\bar{d}_c$) of the W-$1944$, N-$500$, and N-$1848$ codes are respectively
$(3.58,7.16)$, $(4.65,6.87)$, and $(3.79,4.69)$.
\begin{table*}[]
\centering
  \caption {Complexity Summary  } \label{Complexity Table}
          \scalebox{1.1}{
          \begin{tabular}{|c|c|c|c|c|c|}
          \hline
                         & \begin{tabular}[c]{@{}c@{}} C2V\\ Propagation \end{tabular} &
                         \begin{tabular}[c]{@{}c@{}} V2C\\ Update\end{tabular} &
                         \begin{tabular}[c]{@{}c@{}} C2V \\Pre-Update \end{tabular} &
                         CI Update &
                         \begin{tabular}[c]{@{}c@{}} C2V Residual and CI \\ Comparisons \end{tabular}
            \\ \hline
            RBP         & \multirow{7}{*}{1} & \multirow{7}{*}{$\bar{d}_v-1$}  &
            \multirow{7}{*}{\begin{tabular}[c]{@{}c@{}} $(\bar{d}_v-1)\times$\\ $(\bar{d}_c-1)$ \end{tabular} }
            & \multirow{6}{*}{0} & $E-1$
            \\ \cline{1-1} \cline{6-6}
            RDRBP       &  &  &  &  &  $E-1$
            \\ \cline{1-1} \cline{6-6}
            SVNF-RBP    &  &  &  &  &  $\bar{d}_v(\bar{d}_c-1)-1$
            \\ \cline{1-1} \cline{6-6}
            DSVNF-RBP   &  &  &  &  &  $\leq \bar{d}_v(\bar{d}_c-1)-1$
            \\ \cline{1-1} \cline{6-6}
            sLMDRBP & &  &  &  &  $(\bar{d}_v-1)(\bar{d}_c-1)-1$
            \\ \cline{1-1} \cline{6-6}
            LMDRBP       &  &  &  &  &  $(\bar{d}_v-1)\bar{d}_c-1$
            \\ \cline{1-1} \cline{5-6}
            LMD-CIRBP     &  &   &   & $(\bar{d}_v-1)(\bar{d}_c-1)$  &  $(\bar{d}_v-1)\bar{d}_c-1$
            \\ \cline{1-1} \cline{5-6}
             CIRBP       &  &  &  &  $(\bar{d}_v-1)(\bar{d}_c-1)$ &  $N + (1-\kappa)(\bar{d}_v-1) + \kappa(E-1)$
            \\  \hline
          \end{tabular}
          }
  \begin{itemize}
    \item[] \hspace{-1cm} \centering\footnotesize{ $N$: total VN number ~~~~$E^{\*}$: total edge number ~~~~$\bar{d}_v$: averaged VN degree ~~~~$\bar{d}_c$: averaged CN degree ~~~~$\kappa: \Pr(D_{n^*}< \gamma)$}
  \end{itemize}
  \squeezeup
\end{table*}
For the sLMDRBP algorithm, a C2V message propagation is followed by $(\bar{d}_v-1)(\bar{d}_c-1)-1$ comparisons
for deciding the next updated VN and the C2V message to be forwarded. For the LMDRBP algorithm, $(\bar{d}_v-1)
\bar{d}_c-1$ comparisons are required after delivering a C2V message, where $(\bar{d}_v-1)(\bar{d}_c-1)-1$ of
them are used for locating the target VN and the rest of them are for deciding the next updated C2V message. In LMD-CIRBP decoding, passing a C2V message is followed by $(\bar{d}_v-1)
(\bar{d}_c-1)$ CI updates and $(\bar{d}_v-1)\bar{d}_c-1$ comparisons for choosing the ensuing targeted VN and
the associated C2V message to be sent.

For the CIRBP algorithm, there are $(\bar{d}_v-1)(\bar{d}_c-1)$ CI updates after the C2V pre-updates. Then,
$N-1$ and one comparisons are respectively used to search for the largest CI ($D_{n^*}$) and check if $D_{n^*}
\geq\gamma$. If $D_{n^*} \geq \gamma$, additional $\bar{d}_v-1$ comparisons are needed for
selecting the candidate CN; otherwise, we follow the original RBP schedule and perform $E-1$ comparisons to
find the C2V message conveying the maximum residual.
Let $\kappa = \Pr(D_{n^*}< \gamma)$, then $\kappa$ is an increasing function of $\gamma$
and on the average we need $N + (1-\kappa)(\bar{d}_v-1) + \kappa(E-1)$ comparisons to select the updated C2V message.
Our simulation results indicate that $\kappa$ varies with the iteration number and is a function of SNR
and the code used. For W-$1944$ code, (the averaged) $\kappa\approx0.75$ for SNR$=1.5$--$1.75$dB;
for N-$500$ code, $\kappa=0.49$ and $0.51$ for SNR$=2$ and $2.2$ dB; for N-$1848$ code, $\kappa=0.68$
and $0.66$ for SNR$=1.1$ and $1.3$ dB.

Table \ref{Complexity Table} shows that compared with the RBP, RDRBP, and SVNF-RBP decoders, the proposed
LMDRBP and sLMDRBP decoders are more computationally efficient for all codes used. The LMD-CIRBP decoder
is the most complicated except for the CIRBP one since it requires extra complexity for CI update. The
later decoder needs to perform global residual comparison with probability $\kappa$.
The numerical results discussed so far indicate that the proposed decoders provide various tradeoffs between
complexity and decoding performance, and the LMD-CIRBP decoder has the best performance-complexity balance,
offering improved performance at the cost of limited complexity increase.

As mentioned in the last section, the CIRBP and LMD-CIRBP algorithms give impressive first-iteration FER performance
and a valid codeword is likely to be obtained within one iteration (i.e., before $E$ C2V message updates), significantly reducing the average decoding complexity.

%% file: Sec6_MultiEdge.tex
\section{Multi-Edge Updating Strategies}\label{section:multiEdge}
The decoding schedules discussed so far all adopt a single-edge updating strategy that passing one C2V message per update.
To reduce the decoding latency, we propose multi-edge CIRBP (ME-CIRBP) and multi-edge LMD-CIRBP (ME-LMD-CIRBP) algorithms
in this section which allow $N_P$ C2V messages to be propagated in parallel per update.
Specifically, our multi-edge strategy determines $N_P$ VNs to be updated and applies the single-edge strategies to each VN. For
implementation efficiency, the number $N_p$ is fixed in each update.

For a CIRBP based decoding, a simple and intuitive method for simultaneously updating $N_P$ VNs is to choose the nodes with the
largest $N_P$ CI values which requires (at most) $(2N-P-1)N_P /2$ real-number comparisons. To further lower the complexity,
we introduce a VN selection method which selects $N_p$ indices from a candidate VN index set $\mathcal{S}$ for simultaneous updates.
The set of the $N_p$ VN indices selected is denoted by $\mathcal{P}$.

\begin{algorithm}
    \caption{A VN Selection Scheme}
    \label{VN_GSM}
    \begin{algorithmic}[1]
    \Require $N_G$: Group Number, $N_P$: Selected VN Number, $\mathcal{S}$: Input Search Set
    \squeezeupAlgo
    \Ensure $\mathcal{P}$: Selected VN Index Set
    \squeezeupAlgo
    \State Initialize $\mathcal{G}_i=\emptyset$ for $i=0,1,\ldots,N_G-1$, $\mathcal{P}=\emptyset$
    \squeezeupAlgo
    \State $\mathcal{G}_i\leftarrow \mathcal{G}_i\cup \{n\}$, where $i=\lfloor D_n\times N_G\rfloor$, for every $n\in\mathcal{S}$
    \squeezeupAlgo
    \State Find $k^*=\max\{k:|\mathcal{Q}(k)|\leq N_p\}$ and let $\mathcal{P} = \mathcal{Q}(k^*)$
    \squeezeupAlgo
    \If{$| \mathcal{P}|<N_P$}
        \squeezeupAlgo
        \State Randomly choose $N_P-|\mathcal{P}|$ elements from $\mathcal{G}_{N_G-k^*-1}$ to form set $\mathcal{G}'_{N_G - k^* - 1}$
        \squeezeupAlgo
        \State Let $\mathcal{P}\leftarrow\mathcal{P}\cup \mathcal{G}'_{N_G-k^*-1}$
    \squeezeupAlgo
    \EndIf
    \\
    \squeezeupAlgo
    \Return $\mathcal{P}$
    \end{algorithmic}
\end{algorithm}

We first partition $\mathcal{S}$ into $N_G$ groups ($\mathcal{G}_{0},\mathcal{G}_{1},\cdots,\mathcal{G}_{N_G-1}$)
according to their CI values: for all
$n\in\mathcal{S}$, we let $\mathcal{G}_i\leftarrow \mathcal{G}_i\cup \{n\}$ if $D_n\in [i/N_G,(i+1)/N_G)$,
where $N_G$ is a predetermined designed
group number.
We then find $k^*=\max\{k:|\mathcal{Q}(k)|\leq N_p\}$, where $\mathcal{Q}(k)\stackrel{def}{=}\bigcup_{j=1}^k\mathcal{G}_{N_G-j}$.
If $|\mathcal{Q}(k^*)|=N_p$, we let $\mathcal{P}=\mathcal{Q}(k^*)$. Otherwise, we randomly select $N_P-|\mathcal{Q}(k^*)|$ elements from $\mathcal{G}_{N_G-k^*-1}$ to form $\mathcal{G}'_{N_G-k^*-1}$ and set $\mathcal{P}=\mathcal{Q}(k^*)\cup \mathcal{G}'_{N_G-k^*-1}$. The
procedure is formally described in \textbf{Algorithm \ref{VN_GSM}}.

Incorporating the above VN selection method into the CIRBP algorithm, we have the ME-CIRBP algorithm which
we refer to as \textbf{Algorithm \ref{MEPD_algo}}. In this multi-edge updating schedule, the VNs whose indices belong to $\mathcal{P}$
are simultaneously updated. For each selected VN, the corresponding incoming C2V message selection and the subsequent message renewal
procedures are the same as those of the CIRBP algorithm.

\begin{algorithm}
    \caption{Multi-Edge CIRBP (ME-CIRBP) Algorithm}
    \label{MEPD_algo}
    \begin{algorithmic}[1]
    \State Initialize all $L_{m\rightarrow n}^{\text{C}}=0$ and all $L_n=L_{n\rightarrow m}^{\text{V}}=2y_n/\sigma^2$
    \squeezeupAlgo
    \State Generate all $\tilde{L}_{m\rightarrow n}^{\text{C}}$ by (\ref{eqn:C2V}) and compute all $R_{m\rightarrow n}^{\text{C}}$
    \squeezeupAlgo
    \State Compute all $\tilde{L}_n$ and $D_n$
    \squeezeupAlgo
    \State Find $\mathcal{P}$ by \textbf{Algorithm \ref{VN_GSM}} $(\text{\bf{input: }}N_G,N_P,\mathcal{O}_N)$
    \label{ALG:MEPDRBP_findVNs}
    \squeezeupAlgo
    \State For all $p \in\mathcal{P}$, perform lines $8$-$13$ (by letting $n^*\leftarrow p$) in \textbf{Algorithm \ref{PDRBP}}  \textit{in parallel}
    \squeezeupAlgo
    \State Go to line \ref{ALG:MEPDRBP_findVNs} if \emph{Stopping Condition} is not satisfied
    \end{algorithmic}
\end{algorithm}

The multi-edge version of the LMD-CIRBP algorithm (\textbf{Algorithm \ref{MEPDMF_algo}}) is similarly structured:
by combining the LMD-CIRBP decoder with \textbf{Algorithm \ref{VN_GSM}}. In this algorithm, the first $N_P$ targeted VNs
are found from $\mathcal{O}_N$. For each $v_p$, $p\in\mathcal{P}$, we simultaneously carry out the key message updating
procedure of the LMD-CIRBP algorithm (i.e., lines $5$--$10$ of \textbf{Algorithm \ref{LMD-PDRBP}}). For every $p\in\mathcal{P}$,
we update its associated C2V residuals and CI values,
and then we find a VN $v_{p'}$ according to line 8 of \textbf{Algorithm \ref{MEPDMF_algo}}
as the next target VN and add $p'$ to the temporary set $\mathcal{P}'$.
In case different $v_p$'s may suggest the same VN $v_{p'}$
so that $|\mathcal{P}'| < N_P$, we execute \textbf{Algorithm \ref{VN_GSM}} to find the remaining $N_P-|\mathcal{P}|$ VNs
from those VNs which do not belong to $\mathcal{P}'$.

\begin{algorithm}
    \caption{Multi-Edge LMD-CIRBP (ME-LMD-CIRBP) Algorithm}
    \label{MEPDMF_algo}
    \begin{algorithmic}[1]
    \State Initialize all $L_{m\rightarrow n}^{\text{C}}=0$ and all $L_n=L_{n\rightarrow m}^{\text{V}}=2y_n/\sigma^2$
    \squeezeupAlgo
    \State Generate all $\tilde{L}_{m\rightarrow n}^{\text{C}}$ by (\ref{eqn:C2V}) and compute all $R_{m\rightarrow n}^{\text{C}}$
    \squeezeupAlgo
    \State Compute all $\tilde{L}_n$ and $D_n$
    \squeezeupAlgo
    \State Find $\mathcal{P}$ by \textbf{Algorithm \ref{VN_GSM}} $(\text{\bf{input: }}N_G,N_P,\mathcal{O}_N)$
    \squeezeupAlgo
    \State For all $p \in\mathcal{P}$, perform lines $5$-$10$ in \textbf{Algorithm \ref{LMD-PDRBP}} (by letting $n^*\leftarrow p$) \emph{in parallel}
    \label{ALG:MELMD-PDRBP_Prop}
    \squeezeupAlgo
    \State Set $\mathcal{P}'=\emptyset$
    \squeezeupAlgo
    \For {every $p\in\mathcal{P}$}
        \squeezeupAlgo
        \State Find $p'=\arg\max_{j}\{D_j~|~j\in\mathcal{U}(m^*, p)\}$ where $m^* = \arg\max_i\{R^C_{i\rightarrow n^*} | i\in\mathcal{M}(p) \}$
        \squeezeupAlgo
        \State Let $\mathcal{P}' \leftarrow \mathcal{P}'\cup p'$
    \squeezeupAlgo
    \EndFor
    \squeezeupAlgo
    \If{$|\mathcal{P}'|<N_P$}
        \squeezeupAlgo
        \State Find $\mathcal{P}$ by \textbf{Algorithm \ref{VN_GSM}} $(\text{\bf{input: }}N_G,(N_P-|\mathcal{P}'|), \mathcal{O}_N\setminus \mathcal{P}')$
    \squeezeupAlgo
    \EndIf
    \squeezeupAlgo
    \State Let $\mathcal{P}\leftarrow \mathcal{P}' \cup \mathcal{P}$
    \squeezeupAlgo
    \State Go to line \ref{ALG:MELMD-PDRBP_Prop} if \emph{Stopping Condition} is not satisfied
    \end{algorithmic}
\end{algorithm}

\begin{figure}
    \centering
    \subfigure[\label{W_MECI_ite_3} FER and BER performance, $I_{\text{max}}=3$]
    {\epsfxsize=3.15in
    \epsffile{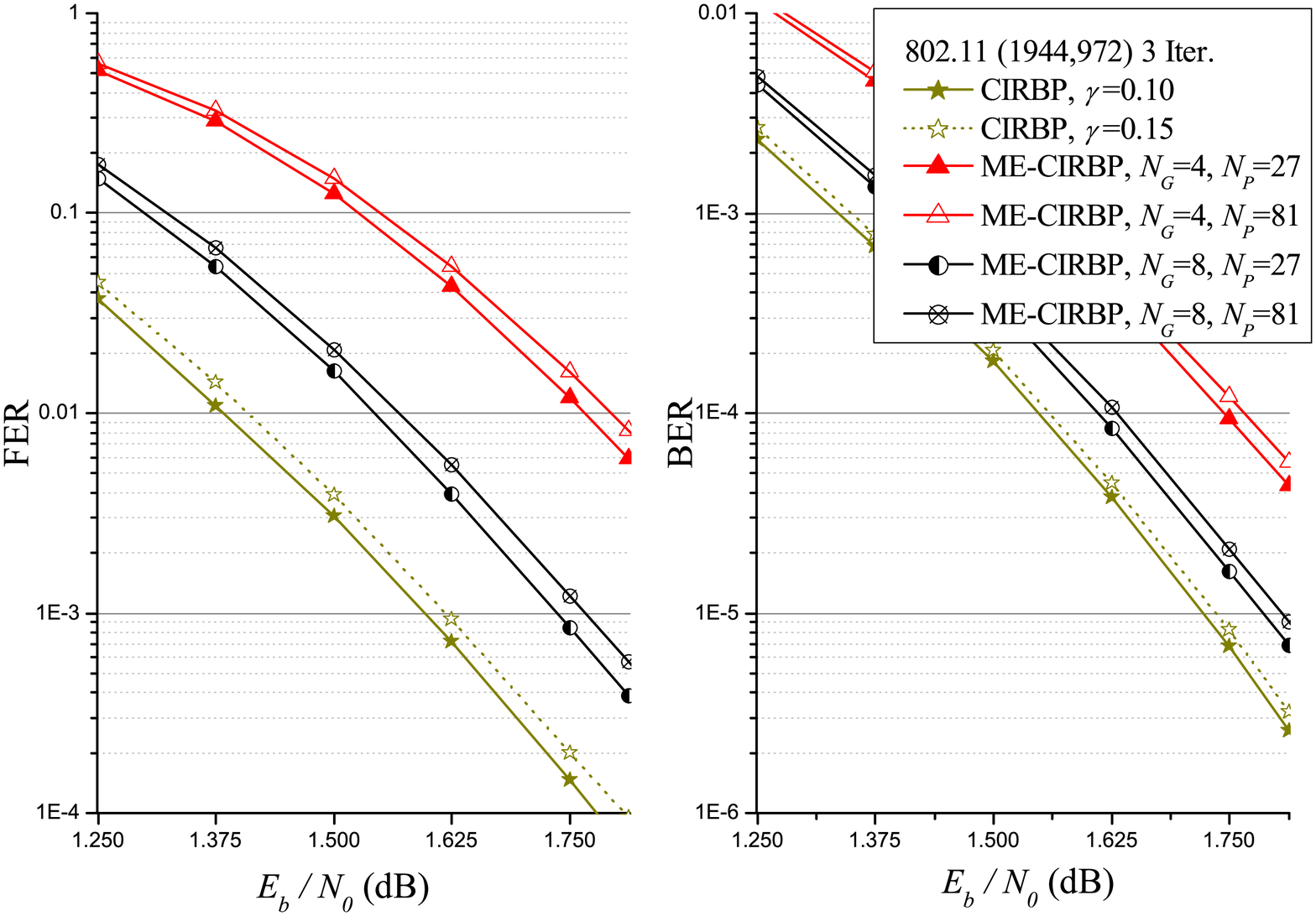}}
    \hspace{0.01\textwidth}
    \subfigure[\label{W_MECI_SNR_175} FER and BER convergence behaviors, SNR = $1.75$ dB]
    {\epsfxsize=3.15in
    \epsffile{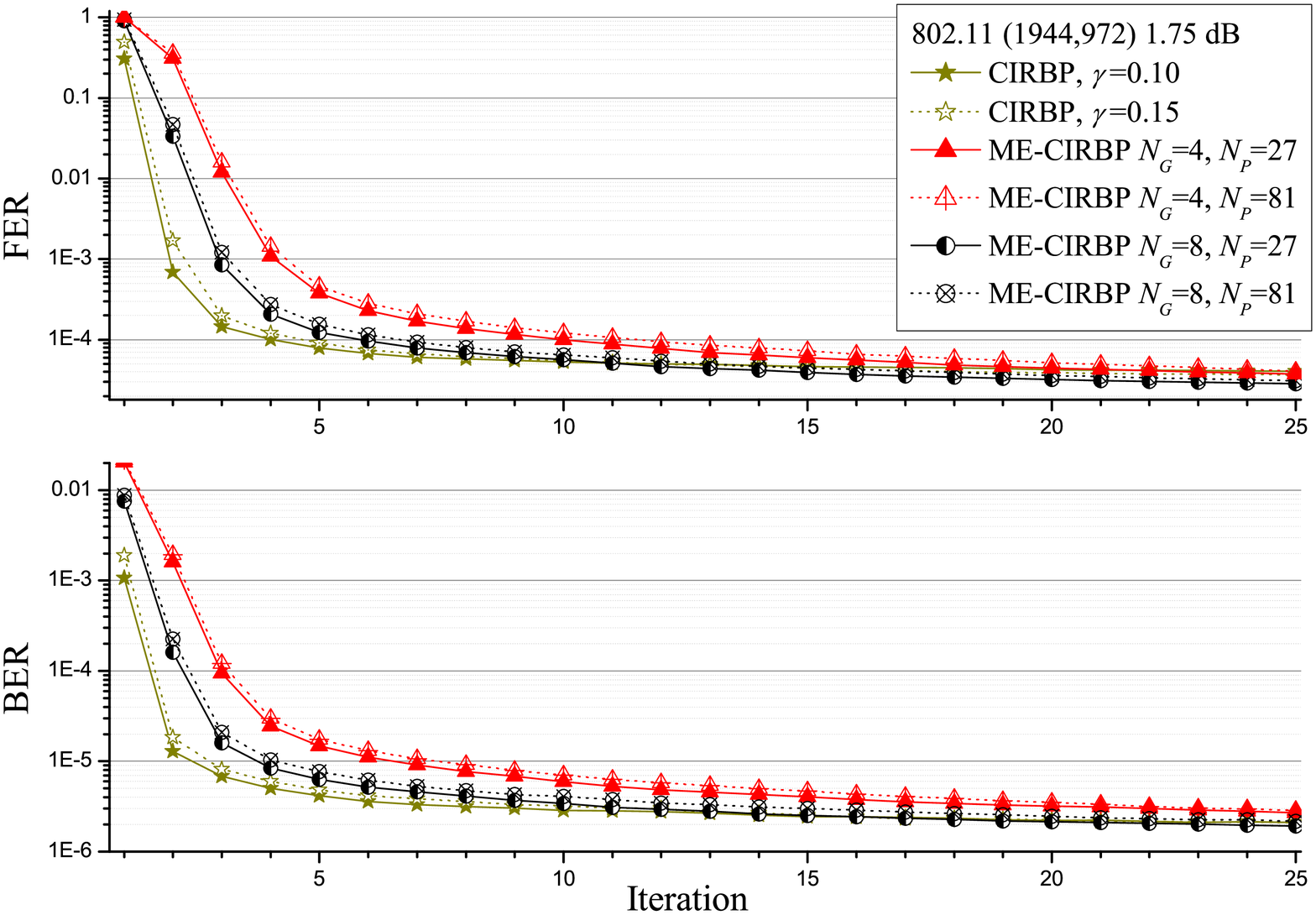}}
    \caption{\label{W_MECI} FER and BER performance of CIRBP and ME-CIRBP algorithms with different $N_P$ and $N_G$ in decoding W-$1944$ code.}
    \squeezeup
\end{figure}

\begin{figure}
    \centering
    \subfigure[\label{W_MELMDCI_ite_3} FER and BER performance, $I_{\text{max}}=3$]
    {\epsfxsize=3.1in
    \epsffile{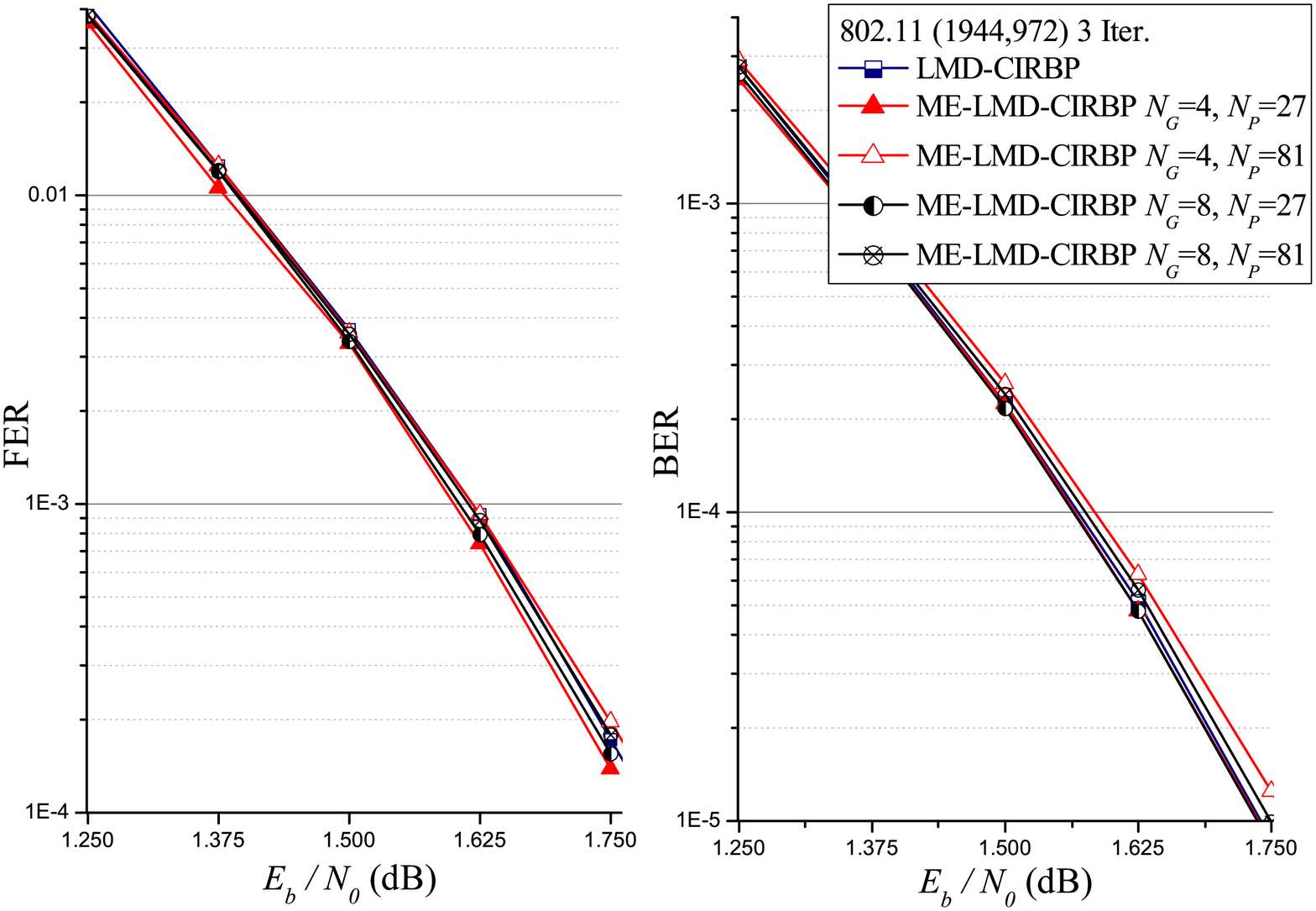}}
     \hspace{0.01\textwidth}
    \subfigure[\label{W_MELMDCI_SNR_175} FER and BER convergence behaviors, SNR = $1.75$ dB]
    {\epsfxsize=3.1in
    \epsffile{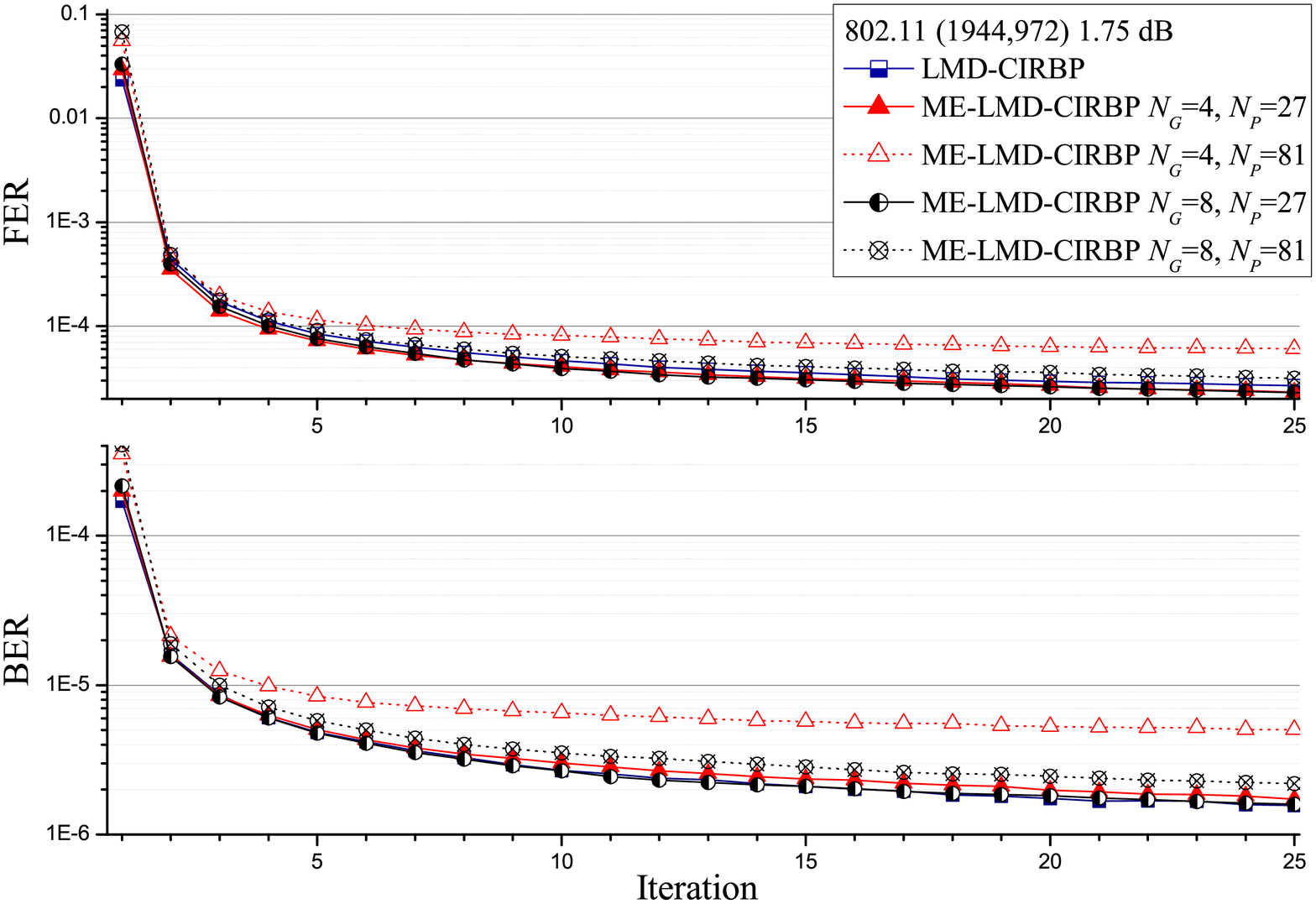}}

    \caption{\label{W_MELMDCI}FER and BER performance of LMD-CIRBP and ME-LMD-CIRBP algorithms with different $N_P$ and $N_G$ in decoding W-$1944$ code.}
    \squeezeup
\end{figure}

\begin{table*}[]
        \centering
          \caption { \label{Table:MEComp} Per-Iteration Complexity of CIRBP, ME-CIRBP, LMD-CIRBP, and ME-LMD-CIRBP Decoders} \label{Complexity Table2}
          \scalebox{0.8}{
          \begin{tabular}{|c|c|c|c|c|c|c|}
          \hline
                         & \begin{tabular}[c]{@{}c@{}} C2V\\ Propagation \end{tabular} &
                         \begin{tabular}[c]{@{}c@{}} V2C\\ Update\end{tabular} &
                         \begin{tabular}[c]{@{}c@{}} C2V \\Pre-Update \end{tabular} &
                         CI Update &
                         \begin{tabular}[c]{@{}c@{}} C2V Residual and CI \\ Comparisons \end{tabular} &

                         \begin{tabular}[c]{@{}c@{}c@{}} Comparisons for\\ Multi-VN Selection\\ (Algorithm 4) \end{tabular}
            \\ \hline
            CIRBP         & \multirow{4}{*}{$E$} & \multirow{4}{*}{\begin{tabular}[c]{@{}c@{}}$E\times$\\$(\bar{d}_v-1)$\end{tabular}}  &
            \multirow{4}{*}{\begin{tabular}[c]{@{}c@{}} $E\times$\\$[(\bar{d}_v-1) (\bar{d}_c-1)]$ \end{tabular} }
            & \multirow{4}{*}{\begin{tabular}[c]{@{}c@{}} $E\times$\\$[(\bar{d}_v-1) (\bar{d}_c-1)]$ \end{tabular} } & $E\times[N + (1-\kappa)(\bar{d}_v-1) + \kappa(E-1)]$ & 0
            \\ \cline{1-1} \cline{6-7}
            ME-CIRBP       &  &  &  &  &  $E\times(d_v-1)$  & $(E/N_P)\times N_G$
            \\ \cline{1-1} \cline{6-7}
            LMD-CIRBP    &  &  &  &  &   $E\times[(\bar{d}_v-1)\bar{d}_c-1]$ &0
            \\ \cline{1-1} \cline{6-7}
            ME-LMD-CIRBP   &  &  &  &  &  $E\times[(\bar{d}_v-1)\bar{d}_c-1]$ & $\leq(E/N_P)\times N_G$

            \\ \hline
          \end{tabular}
          }
      \begin{itemize}
        \item[] \hspace{-1cm} \centering\scriptsize{ $N$: total VN number ~~$E^{\*}$: total edge number ~~$\bar{d}_v ~(\bar{d}_c)$: averaged VN (CN) degree ~~ $N_G$: group number ~~ $N_P$: selected VN number }
      \end{itemize}
\end{table*}
We plot the performance and convergence behaviors of the ME-CIRBP and ME-LMD-CIRBP algorithms and their single-edge versions
in decoding the W-$1944$ code in Figs. \ref{W_MECI} and \ref{W_MELMDCI}. The channel and modulation scheme are the same as
those specified in Sec. \ref{section:simulation}. Fig. \ref{W_MECI} shows that the ME-CIRBP algorithm suffers from performance
loss at early decoding iterations (but requires only $1/N_P$ decoding latency). As expected, the error-rate performance of
both ME decoders improves with a larger $N_G$ or a smaller $N_P$. Fig. \ref{W_MELMDCI_SNR_175} demonstrates that, except for
the case $(N_G, N_p)=(4,81)$ and at the very first iteration, the ME-LMD-CIRBP algorithm provides BER and FER performance
comparable to that of its single-edge version. Both figures show that with a judicial choice of $(N_G, N_P)$, the proposed
ME algorithms yield similar or even better converged performance and, under a low latency constraint, they give far better FER
performance.

In Table \ref{Table:MEComp}, we compare the per iteration complexities of the CIRBP, ME-CIRBP, LMD-CIRBP, and ME-LMD-CIRBP decoders.
For ME-CIRBP decoder (\textbf{Algorithm 5}), $N_P$ VNs are selected by \textbf{Algorithm 4} and then updated. This select-VN-then-update procedure repeats $E/N_P$ times in one iteration (and propagate $E$ C2V messages in total).
We assume that the VN grouping in \textbf{Algorithm 4} (line 2) can be simply performed by assigning $n$ to $\mathcal{G}_{\lfloor D_n\times N_G\rfloor}$ or equivalently by passing $D_n$ through an $N_G$-level uniform quantizer. Hence, executing \textbf{Algorithm 4} once requires
at most $N_G$ integer comparisons where (at most) $N_G-1$ of them are for finding $k^*$ (line 3) and the remaining ones are for checking if $|\mathcal{P}|<N_P$ (line 4). The ME-CIRBP thus requires $(E/N_P)\times N_G$ integer comparisons for the VN selection in each iteration. As the ME-CIRBP decoder need not compare CI after VN selection, it consumes only $E\times (\bar{d}_v-1)$ real-value comparisons for comparing the C2V residuals of the selected VNs per iteration. The remaining operations are the same as the CIRBP decoder. As summarized in Table \ref{Table:MEComp}, when $N\times N_P>N_G$, the ME-CIRBP decoder requires less computational efforts compared with the CIRBP decoder.

The complexity associated with the ME-LMD-CIRBP decoder can be similarly evaluated. As $E$ C2V messages will be propagated in one iteration,
the per-iteration complexity required for updating messages/CIs and residual comparisons in the ME-LMD-CIRBP decoding (lines 5-10 of \textbf{Algorithm 6}) is the same as that needed by the LMD-CIRBP decoder. However, because \textbf{Algorithm 4} is executed at most $E/N_P$ times in an iteration for the case {$|\mathcal{P}'|<N_P$} occurs in ME-LMD-CIRBP decoding (lines 11-13 of \textbf{Algorithm 6}), compared with the LMD-CIRBP decoder, the ME-LMD-CIRBP decoder may consume at most extra $(E/N_P)\times N_G$ integer comparisons per iteration.
To summarize, the ME-CIRBP algorithm generally consumes less computational effort compared with the CIRBP decoder but suffers from greater performance loss; the ME-LMD-CIRBP decoder may offer quite-nice performance-latency tradeoffs at the cost of slightly increased complexity.

%% file: Sec7.tex
\section{Conclusion}\label{section:conclusion}
In this paper, we have presented novel IDS LDPC decoding schedules which apply a VN selecting metric called conditional innovation
and a search complexity reduction criterion that limits our target VN/CN search range to those newly updated CNs and their connected
VNs. The proposed schedules are VN-centric in the sense that the metrics used are aimed to improve the reliability of the target VNs'
bit decisions by predicting the probability of reversing potential incorrect decisions. Computer simulation results indicate that
our schedules outperform known schedules and achieve most impressive error rate performance gain in the first few iterations.
Therefore, as far as the average computing complexity is concerned, the proposed schedules do not incur more computing burden. The
converged FER performance of the LMD-based algorithms against their counterparts indicates that the search range reduction will
eventually include those VNs that should be updated. The outstanding first-iteration performance of the LMD-CIRBP algorithm may be
attributed to the decreasing probability of improper update selections by considering only the shortlist candidates. To shorten the
decoding delay, we develop multi-edge versions of the CIRBP and LMD-CIRBP algorithms by increasing the degrees of parallelism in
updating. The multi-edge versions are of low latency and are proved to be efficient in performance.

%% file: Appendix_0214.tex
\linespread{1.5}
\section{A Semi-analytic Proof of {\it Property 1}}\label{app:A}
We verify Property~1 by evaluating (\ref{obj_func}) using the GA-DE technique \cite{DEol}.
Recall that $D=|\tilde{P_0}-P_0|$, where $0\leq \tilde{P_0}, P_0 < 1$.
Conditioning on $D\geq \gamma$, the numerator of (\ref{obj_func}) is equal to
\begin{eqnarray}
        \nonumber
        \Pr \left( P_0\geq 0.5 | D \geq \gamma \right)
        &=& \Pr \left( P_0\geq \max(\gamma, 0.5) | D \geq \gamma \right)
        ~=~ \int_{\max(\gamma, 0.5)}^{1} f_{P_0 | D}(\tau | D \geq \gamma ) \,\mathrm{d}\tau \\
        \nonumber
        &=& \int_{\max(\gamma, 0.5)}^{1} \frac{\Pr ( D \geq \gamma | P_0=\tau ) f_{P_0}( \tau )}{\Pr ( D \geq \gamma) } \,\mathrm{d}\tau,
\end{eqnarray}
since $P_0-\gamma \geq \tilde{P}_0 \geq 0$, where $f(\cdot)$ stands for probability density function (PDF); similarly, as $P_0 + \gamma \leq \tilde{P_0} \leq 1$, the denominator of (\ref{obj_func}) is equal to
\begin{eqnarray} \nonumber
         \Pr \left( P_0<0.5 | D \geq \gamma \right)=
         \int_{0}^{\min(1-\gamma, 0.5)} \frac{\Pr ( D \geq \gamma | P_0=\tau ) f_{P_0}(\tau)}{\Pr ( D \geq \gamma )} \,\mathrm{d}\tau.
\end{eqnarray}
Combining the above expressions then yields that
\begin{eqnarray}\label{appA_func2}
    \mathcal{J}(\gamma)
    =  \frac{\int_{\max(\gamma, 0.5)}^{1} \Pr ( D \geq \gamma | P_0=\tau ) f_{P_0}(\tau) \,\mathrm{d}\tau }  {\int_{0}^{\min(1-\gamma, 0.5)}
    \Pr(D \geq \gamma | P_0=\tau ) f_{P_0}(\tau) \,\mathrm{d}\tau }.
\end{eqnarray}

We now apply the GA-DE to obtain $f_{P_0}(\tau)$ and $\Pr( D \geq \gamma | P_0=\tau)$.
Note that in the GA-DE, all messages are modeled as i.i.d. consistent Gaussian random variables; specifically, the C2V (resp. V2C) messages are distributed according to $\mathcal{N}(\mu_C, 2\mu_C)$ (resp. $\mathcal{N}(\mu_V, 2\mu_V)$), where $\mu_C$ (resp. $\mu_V$) denotes the mean of the C2V (resp. V2C) messages.
Due to the all-zero codeword assumption, the mean of the LLR of the received signal is $\mu_0=2/\sigma^2$ and hence we initialize $\mu_{V} = \mu_0$.
For $(d_v, d_c)$ regular LDPC codes, the $\mu_{C}$ and $\mu_{V}$ are recursively calculated by (we have dropped the iteration index for notational simplicity):
\begin{eqnarray}\label{DE-algo}
  \mu_{C} &=& \Phi^{-1}\left( 1-\left[ 1-\Phi\left( \mu_{V} \right) \right]^{d_c-1} \right), \\
  \mu_{V} &=& \mu_0 + (d_v-1)\mu_{C}
\end{eqnarray}
where $\Phi(\mu)$ is given in \cite[Definition 1]{DEol}.
Similar recursions for irregular LDPC codes can be found in \cite{DEol}.

Following the idea of the GA-DE, we approximate the total LLR $L$ and the precomputed total LLR $\tilde{L}$ as consistent Gaussian random variables, i.e., $L \sim \mathcal{N}(\mu_{L},2\mu_{L})$ and $\tilde{L}\sim \mathcal{N}(\mu_{\tilde{L}}, 2\mu_{\tilde{L}})$, where $\mu_{L}= \mu_0 + d_v\mu_{C}$.
Moreover, their difference $\Delta L \triangleq \tilde{L} - L$ is also approximated in the same way with mean $\mu_{\Delta L}= \mu_{\tilde{L}} - \mu_{L}$, i.e., $\Delta L \sim\mathcal{N}(\mu_{\Delta L}, 2\mu_{\Delta L})$.
Using the above approximations and the definitions $L= \ln (P_0 / P_1) $ and $Q(\alpha)=\frac{1}{\sqrt{2\pi}} \int_{\alpha}^{\infty} e^{-\frac{\beta^2}{2}} \mathrm{d}\beta$, we obtain
\begin{eqnarray}
    \label{eqn:pdfpn0}
    f_{P_0}(\tau)=f_{L}\left(\ln\left(\frac{\tau}{1-\tau}\right)\right),
\end{eqnarray}
and
\begin{eqnarray}
    \nonumber
    \Pr(D \geq \gamma | P_0=\tau ) &=&\Pr \left (\tilde{P}_0 \geq \min( \tau +\gamma, 1) ~\mathrm{or}~ \tilde{P}_0 \leq \max(\tau - \gamma, 0) | P_0=\tau \right )
    \\ \nonumber
    &=& 1 - \int_{ \max(\tau-\gamma, 0) }^{ \min(\tau+\gamma,1) } f_{\tilde{L}|L} \left( \ln \left(\frac{\tilde{\tau}}{1-\tilde{\tau}} \right) \bigg| \ln \left(\frac{\tau}{1-\tau} \right) \right) \,\mathrm{d}\tilde{\tau} 
    \\ \nonumber
    &=&1 - \int_{ \max(\tau-\gamma, 0) }^{ \min(\tau+\gamma,1) }f_{\Delta{L}} \left( \ln \left(\frac{\tilde{\tau}}{1-\tilde{\tau}} \right)-\ln \left(\frac{\tau}{1-\tau} \right) \right) ~ \mathrm{d}\tilde{\tau}
    \\ \nonumber
    &=& 1-
    Q\left( \frac{ \ln\left(\frac{\max(\tau-\gamma, 0)}{1-\max(\tau-\gamma, 0)}\right) - \ln\left(\frac{\tau}{1-\tau}\right) - \mu_{\Delta L}}{\sqrt{2\mu_{\Delta L}}} \right)
    \\ \label{eqn:pdfdeltaln}
    & &\qquad +
    Q\left( \frac{ \ln\left(\frac{\min(\tau+\gamma,1)}{1-\min(\tau+\gamma,1)}\right) - \ln\left(\frac{\tau}{1-\tau}\right)
    - \mu_{\Delta L}}{\sqrt{2\mu_{\Delta L}}} \right).
\end{eqnarray}

Given $\mu_{L}$ and $\mu_{\tilde{L}}$ obtained from (\ref{DE-algo}) for any fixed iteration, we can calculate $\mathcal{J}(\gamma)$ as a function of $\gamma$ using (\ref{eqn:pdfdeltaln}), (\ref{eqn:pdfpn0}), and (\ref{appA_func2}).
The GA-DE curves in Fig. \ref{PD_plot_GA8000} are the $\mathcal{J}(\gamma)$'s for the first three iterations with the flooding schedule.
The curves almost coincide with the simulated ones, and the decreasing property of $\mathcal{J}(\gamma)$ as claimed in Property~1 is also revealed. We remark that similar behavior is observed for other LDPC codes of different rates and degree distributions. Moreover, our
proof relies only on the assumption that $\mu_{\Delta L} > 0$ whence is independent of the BP-based schedule used.

\section{Proof of {\it Property 2}}\label{app:B}
\setcounter{equation}{0}
We prove that $F(\gamma)>1$ by considering two cases: $\gamma \geq P_0$ and $\gamma < P_0$.
Note that the event $\{D\ge\gamma\}$ implies that $\tilde{P}_0$ can lie in $[0, P_0 - \gamma]$ or $[P_0 + \gamma, 1)$.
When $P_0<0.5$ and $\gamma \geq P_0$, we must have that $\tilde{P}_0\in[P_0 + \gamma, 1)$ and hence $\tilde{P}_0\ge P_0+\gamma$ with probability $1$, resulting in that $F(\gamma)=\infty$.
For the case $\gamma< P_0$, we first rewrite (\ref{app_eq1}) as 
\begin{eqnarray}
  F(\gamma)= \frac{\Pr (\{\tilde{P}_0\ge P_0\} \cap \{D\ge\gamma\})}
  { \Pr(\{\tilde{P}_0<P_0\} \cap \{D\ge\gamma\})}=\frac{\Pr (\tilde{P}_0 \geq P_0+\gamma )}
  { \Pr (\tilde{P}_0  \leq P_0-\gamma ) }
  = \frac{ \int_{P_0+\gamma}^{1} f_{\tilde{P}_0}(\tilde{\tau}) \,\mathrm{d}\tilde{\tau} }
  { \int_{0}^{P_0-\gamma} f_{\tilde{P}_0}(\tilde{\tau}) \,\mathrm{d}\tilde{\tau} }. \label{app_eq1.5}
\end{eqnarray}
Since $f_{\tilde{P}_0}(\tilde{\tau} ) = f_{\tilde{L}}(\ln ( \tilde{\tau} /(1-\tilde{\tau} )))$ and $\tilde{L}\sim \mathcal{N}(\mu_{\tilde{L}},2\mu_{\tilde{L}})$, we have the following expressions
\begin{eqnarray}\nonumber
  \int_{P_0+\gamma}^{1} f_{\tilde{P}_0}(\tilde{\tau}) \,\mathrm{d}\tilde{\tau}= Q\left( g_1(P_0,\gamma) \right)\ \text{and}\
~~\int_{0}^{P_0-\gamma} f_{\tilde{P}_0}(\tilde{\tau}) \,\mathrm{d}\tilde{\tau}= Q\left( g_2(P_0,\gamma) \right)
\end{eqnarray}
for the terms in \eqref{app_eq1.5}, where $Q(\cdot)$ is defined in Appendix~A and
\begin{eqnarray} \nonumber
    g_1(P_0,\gamma) = \frac{\ln\left( \frac{P_0+\gamma}{1-(P_0+\gamma)} \right) -\mu_{\tilde{L}}}{\sqrt{2\mu_{\tilde{L}}}},
~~
    g_2(P_0,\gamma)= \frac{\mu_{\tilde{L}} - \ln\left( \frac{P_0-\gamma}{1-(P_0-\gamma)} \right) }{\sqrt{2\mu_{\tilde{L}}}}.
\end{eqnarray}
With the above quantites, the expression in (\ref{app_eq1.5}) is simplified as
\begin{eqnarray}
F(\gamma) = \frac{Q\left( g_1(P_0,\gamma) \right) }{Q\left( g_2(P_0,\gamma) \right) }.
\end{eqnarray}
Since $P_0< 0.5$, we obtain that
\begin{eqnarray} \nonumber
  \left[\ln\left( \frac{P_0+\gamma}{1-(P_0+\gamma)} \right)-\mu_{\tilde{L}} \right]< \left[\mu_{\tilde{L}}-\ln\left( \frac{P_0-\gamma}{1-(P_0-\gamma)} \right)\right],
\end{eqnarray}
which implies that $Q(g_1(P_0,\gamma)) >  Q(g_2(P_0,\gamma))$ and hence $F(\gamma)>1$.

Based on the above derivation, it is clear that $F(\gamma)=\infty$ for $\gamma\ge P_0$. We next show that $F(\gamma)$ is strictly increasing for $\gamma\in[0, P_0)$.
Specifically, we prove the following derivative is positive.
\begin{eqnarray}  \nonumber
  \frac{\mathrm{d} F(\gamma)}{\mathrm{d} \gamma}
  &=&
  \frac{1}{\sqrt{2\mu_{\tilde{L}}}\left(Q\left(g_2(P_0, \gamma) \right)\right)^2}
  \\ \label{app_eq2.5}
  & &
  \times \left[ \frac{Q'\left(g_1(P_0, \gamma) \right) Q\left(g_2(P_0, \gamma) \right)}{(P_0+\gamma)(1-(P_0+\gamma))}
  - \frac{Q'\left(g_2(P_0, \gamma) \right) Q\left(g_1(P_0, \gamma) \right)}{(P_0-\gamma)(1-(P_0-\gamma))} \right]
\end{eqnarray}
where
\begin{eqnarray}
\nonumber
Q'(\alpha) \triangleq \frac{\mathrm{d} Q (\alpha)}{\mathrm{d} \alpha}= \frac{-\exp(-\alpha^2/2)}{ \sqrt{2\pi}}.
\end{eqnarray}
Recall the facts that $Q(\alpha)>0$, $Q'(\alpha)<0~\forall\ \alpha \in \mathds{R}$, and $\mathrm{d} Q'(\alpha) / \mathrm{d} \alpha = -\alpha Q'(\alpha)$. Defining $P(\alpha) \triangleq Q(\alpha)/Q'(\alpha)$, one can show that $P'(\alpha) \triangleq \mathrm{d} P(\alpha)/\mathrm{d} \alpha
= [(Q'(\alpha))^2 + \alpha Q(\alpha)Q'(\alpha)] / \left(Q'(\alpha) \right)^2 \allowbreak >0$ for $\alpha\le 0$.
For $\alpha>0$, we apply the inequality $\alpha Q(\alpha)< -Q'(\alpha)$
\cite{Q_upper} to conclude that $Q'(\alpha) (Q'(\alpha) + \alpha Q(\alpha)) > 0$.
Since $P'(\alpha) > 0$ for all $\alpha$, i.e., $P(\alpha)$ is increasing, and $g_2(P_0,\gamma)>g_1(P_0,\gamma)$, we have that
\begin{eqnarray}
\label{app_eq3}
  P(g_2(P_0,\gamma)) > P(g_1(P_0,\gamma)).
\end{eqnarray}
Using (\ref{app_eq3}) and the fact that $(P_0+\gamma)(1-(P_0+\gamma)) > (P_0-\gamma)(1-(P_0-\gamma))$,
we further obtain
\begin{eqnarray}
\label{app_eq4}
  \frac{Q'\left(g_1(P_0,\gamma) \right) Q\left(g_2(P_0,\gamma) \right)}{(P_0+\gamma)(1-(P_0+\gamma))}
  >
  \frac{Q'\left(g_2(P_0,\gamma) \right) Q\left(g_1(P_0,\gamma) \right)}{(P_0-\gamma)(1-(P_0-\gamma))}.
\end{eqnarray}
Substituting (\ref{app_eq4}) into (\ref{app_eq2.5}) then shows that $\mathrm{d} F(\gamma) / \mathrm{d} \gamma>0$ for $\gamma\in[0, P_0)$.